\documentclass[12pt,useAMS,usenatbib]{article} 
\usepackage{graphics}
\usepackage{graphicx} 
\usepackage{amsmath} 
\usepackage{jcappub}
\usepackage{natbib} 
\usepackage{hyperref} 
\usepackage{color}
\usepackage{psfrag} 
\usepackage{epsfig}
\usepackage{comment}
\usepackage[utf8x]{inputenc}

\title{\boldmath Using the redshift evolution of the Lyman-$\alpha$ effective opacity as a
probe of dark matter models }

\author[a]{Anjan Kumar Sarkar}
\author[b]{Kanhaiya L. Pandey}
\author[a]{Shiv K. Sethi}
\affiliation[a]{Raman Research Institute, Sadashivnagar, Bangalore, Karnataka 560080}
\affiliation[b]{International Center for Theoretical Sciences, TIFR, Shivakote, Hesaraghatta, Bangalore, Karnataka 560089}
\emailAdd{anjans@rri.res.in}
\emailAdd{kanhaiya.pandey@icts.res.in}
\emailAdd{sethi@rri.res.in}

\abstract{ Lyman-$\alpha$ forest data are  known to be a good probe of the small scale matter power. In this paper, we explore the redshift evolution of the observable  effective optical depth $\tau_{\rm eff} (z)$ from the Lyman-$\alpha$ data as a discriminator between dark matter models that differ from the $\Lambda$CDM model on small scales. We consider the 
  thermal warm dark matter (WDM) and the ultra-light axion (ULA) models for the following set of parameters: the mass of ULA, $m_a \simeq 10^{-24}\hbox{--}5 \times 10^{-22} \, \rm eV$ and WDM mass, $m_{\rm wdm} = 0.1 \hbox{--} 4.6 \, \rm keV$. We simulate the line-of-sight HI density and velocity fields using semi-analytic methods. The  simulated effective optical depth for the  alternative dark matter models  diverges from the $\Lambda$CDM model for  $z \gtrsim 3$, which provides a meaningful probe of the matter power at small scales. Using likelihood analysis, we compare the simulated data with the high-resolution Lyman-$\alpha$ forest data in the
  redshift range $2 < z < 4.2$. The analysis yields the following 1$\sigma$ bounds on dark matter masses: $m_{\rm wdm} > 0.7\, {\rm keV}$ and $m_{\rm a} > 2 \times 10^{-23} \, {\rm eV}$. To further test the efficacy of
 our proposed method, we simulate synthetic data sets compatible with the $\Lambda$CDM
model in the redshift range $2 \leq z \leq 6.5$ and compare with theory. The 1$\sigma$ bounds obtained 
are significantly tighter: $m_{\rm wdm} > 1.5 \, {\rm keV}$ and $m_{\rm a} > 7 \times 10^{-23} \, {\rm eV}$. 
Although our
method provides an alternative way of constraining dark matter models, we note that
these bounds are weaker than those obtained by  high-resolution hydrodynamical simulations. }

\begin{document}
\maketitle
\flushbottom

\section{Introduction}

The standard cosmological model has proved to be spectacularly successful
during  the past three decade. Among other probes, the  measurement of CMB temperature and polarization
anisotropies,  galaxy clustering as revealed by large surveys, and the detection of high-redshift supernova~1a  have been
key to this success \citep{hinshaw2013nine,Sievers:2013ica,aghanim2018planck,abolfathi2018fourteenth,hicken2009improved,conley2010supernova,ho2012clustering}. An important ingradient of the concordance $\Lambda$CDM  model is the cold dark matter. However, even after extensitve  laboratory and astronomical searches, the nature of dark matter is yet to be directly determined. Its properties are
indirectly inferred based on many  observations covering a wide range of 
 length scales and epochs of the universe 
(e.g. \cite{aghanim2018planck,Sievers:2013ica,abolfathi2018fourteenth,Bartelmann:1999yn}).

 In the $\Lambda$CDM model, the cold dark matter particle corresponds to the Weakly Interacting Massive Particle (WIMP), which in turn is inspired by the 
 well-known WIMP miracle\cite{Craig:2015xla}. The supersymmetric extension of the standard model of particle physics is consistent with  a particle with self-annihilation cross-section $\langle \sigma v\rangle \sim 3 \times 10^{-26}\rm{cm^3s^{-1}}$ and mass in the range 100--1000 GeV. This theory  correctly  predicts the  dark matter abundance inferred by the cosmological observations. This
 coincidence has spurred many direct 
\cite{Angloher:2011uu,Aprile:2010um,Ahmed:2010wy,Akerib:2013tjd}, indirect\cite{Adriani:2010rc,FermiLAT:2011ab,Aguilar:2007yf} and 
collider \cite{Goodman:2010yf,Fox:2011pm} searches  of the WIMP worldwide. However,  none of these experiments have yet succeeded  in providing 
consistent information about the particle nature of the  dark matter.

While CMB and galaxy clustering observations show that the  CDM is a good
candidate of dark matter for scales $k < 0.1 \, \rm Mpc^{-1}$,  there exist  long-standing astrophysical  issue with the model at smaller scales. N-body simulations based on the CDM model  predict  an order of magnitude larger number of satellite galaxy of
the Milky way as compared to the observations \citep{moore1999dark,klypin1999missing,peebles2010nearby,diemand2007formation}. CDM N-body simulations predict
a cuspy profile at the center of galaxies but the observed profile is flat \citep{navarro1997universal,stadel2009quantifying}.   Another issue  to emerge from the comparison of N-body simulation with observations  is the  ``too big to fail'' problem 
\cite{Garrison-Kimmel:2014vqa,BoylanKolchin:2011de}.  All these issues provide a  motivation  to go beyond the standard CDM  paradigm  and to  consider alternatives,  which differ from the CDM on galactic scales but  reproduce its success on 
the cosmological scales.

The Lyman-$\alpha$ forest are the  observed  absorption features along the directions of QSOs owing to  the fluctuations in the neutral hydrogen (HI) density of the predominantly ionized diffuse IGM in the post-reionization era ($z \leq 6$).  Hydrodynamical simulations have shown these fluctuations correspond to
mildly non-linear density contrast ($\delta < 10$) of the underlying density
field (e.g. \cite{croft1998,croft1999,croft2002,mcdonald,rauch1998lyman} and references therein). As this allows one to probe the fluctuations of the density field
at scales comparable to  the Jeans' scale of the IGM ($k \simeq 5\hbox{--}7 \rm Mpc^{-1}$) in the redshift range $2 < z < 5$, the Lyman-$\alpha$ forest provide a suitable setting for   the measurement of the matter power spectrum for a  vast range of  scales including small scales not accessible to other probes  such as the galaxy surveys\citep{croft1998,croft1999,croft2002,mcdonald,rauch1998lyman}. These data have found  widespread   applications in cosmology, e.g. the measurement of  bispectrum \citep{mandelbaum2003,viel2004}, the estimation of the cosmological parameters \citep{hernquist1996lyman,mcdonald1999,lesgourgues2007}, obtaining constraints on the neutrino mass \citep{croft1999cosmological,yeche2017}, dark energy \citep{mcdonald2007dark}, and the detection  of the baryon Acoustic Oscillations (BAO) \citep{slosar2013measurement,delubac2013baryon,delubac2015baryon}.

In this paper we propose a novel method to investigate alternative
dark matter models using Lyman-$\alpha$ forest data. These  data  allow us to  measure the  flux decrement as a function of redshift, which can be  quantified  in terms of the Lyman-$\alpha$ effective optical depth $\tau_{\rm eff}(z)$. The observational data have determined  $\tau_{\rm eff}$ over the redshift range $2 \leq z \leq 5$ 
for a range of spectral resolutions \citep{faucher2008,bolton2009evolution,becker2013,kamble2020measurements}. For comparison with the data, we simulate the one-dimensional HI density
and velocity fields using semi-analytic methods (e.g. \cite{bidavidson1997})
 to compute  the effective optical depth in the redshift range of interest. 
 The redshift evolution of the theoretically estimated effective optical depth
 is compared with  the 
Lyman-$\alpha$ data. We  explore whether this method can be used to discriminate 
 between dark matter models with different  small-scale matter power
 (e.g. \cite{2013ApJ...762...15P}). We also compare theoretical predictions
 with simulated data.

We study  two alternative dark matter models for comparison with 
 the Lyman-$\alpha$ data. One of these  models is the  warm dark matter (WDM) model, which has been extensively studied for cosmological applications \citep{zhang2009galactic,boyarsky2008constraints,boyarsky2009realistic,seljak2006can,smith2011}.  In the model we consider, the dark matter particle
is coupled to the thermal bath in the early universe and its mass lies in the
range from a few hundred eVs to 10~keV. Thermally-produced gravitinos and the sterile neutrino are the few possible candidates for the WDM particles \citep{ellis1984supersymmetric,dodelson1994sterile}. The ultra-light axions (ULA) arise
naturally within the framework of axiverse \citep{arvanitaki2010string}. The ULA
have masses in the range $10^{-33} < m_a < 10^{-20} \, \rm eV$ and behave
like a coherent scalar field. For both the WDM and the ULA  models, the matter power is suppressed at small scales and these models have been studied for various astrophysical and cosmological applications  (\cite{frieman1995cosmology,coble1997dynamical,hu2000fuzzy,marsh2010ultralight,park2012axion,kobayashi2017lyman,viel2013warm,Polisensky:2010rw, Anderhalden:2012qt, Lovell:2011rd,Maccio:2012qf, Schneider:2011yu,Baur:2015jsy, Marsh:2015wka, Marsh:2015xka,Hui:2016ltb,2016JCAP...04..012S,2017JCAP...07..012S,smith2011,hlozek2015,rogers2020strong}).
These models have also been studied using Lyman-$\alpha$ forest data
for constraining small-scale power 
(\cite{palanque2013one,irvsivc2017lyman,chabanier2019one,viel2013warm,kobayashi2017lyman}). 
We also discuss
how our results compare with these findings.

In the following section, we briefly discuss the two alternative dark matter models 
we study. In section~\ref{sec:simudet}, the semi-analytic  simulations are described 
in detail. In section~\ref{sec:obsqua} we discuss the observables 
and the data. 
In section~\ref{sec:resu}, we present our main results. 
In section~\ref{sec:summ}, we summarize the main findings 
and conclude. 

Throughout this paper, we  use the following cosmological parameters: $\Omega_{c 0} = 0.2285$, $\Omega_{b 0} = 0.046$, $\Omega_{\nu 0} = 0.0013$, $\Omega_{k 0} = 0$, $\Omega_{\Lambda 0} = 0.7242$ and $h = 0.7$ \citep{aghanim2018planck}.

\section{Alternative dark matter models: warm dark matter and ultra-light axions} \label{sec:dmmod}

In this section, we  briefly discuss the two alternative dark matter models
we consider in our study. They are both motivated by the observational evidence
that while the usual $\Lambda$CDM model is in excellent agreement
with CMB and galaxy clustering at at $k \lesssim 0.2 \, \rm Mpc^{-1}$, it overpredicts the matter power at smaller
scales. Both the warm dark matter (WDM)  the ultra-light axion (ULA) models
yield diminished matter power at small scales. 

In the WDM scenario we consider, the dark matter particles  are lighter
than the cold dark matter particles and are coupled to the thermal
bath at early times.  When the  temperature of the universe is much
larger than the mass of the dark matter particle, $T \gtrsim m_{\rm wdm}$,  the particles are 
relativistic and  free-stream, which causes  suppression of the matter power
at scale smaller than horizon size corresponding to the time at which
$T\simeq m_{\rm wdm}$(e.g. \citep{bond1980massive,viel2005constraining,smith2011testing,seigar2015cold,kang2020warm}. 
At later times, the dark matter particles 
become  non-relativistic and their velocity dispersion falls as $1/a$, which causes them to  behave as the  CDM particles at late times.  The   free-streaming length scale $k_{\rm fs}$ is given by \citep{zentner2003halo,bode2001halo}:
\begin{equation}
  k_{\rm fs} \simeq \left(\frac{0.3}{\Omega_{\rm wdm}} \right)^{0.15} \,\, \left(\frac{m_{\rm wdm}}{\rm{keV}} \right)^{1.15} \rm {Mpc^{-1}}
\end{equation}
where $\Omega_{\rm wdm}$ gives the WDM matter density expressed in the 
units of the critical  
mass density of the universe.  For 
$m_{\rm wdm} = 0.3 \, {\rm keV}$, the free-streaming length scale, $k_{\rm fs} \simeq 0.3\, {\rm Mpc^{-1}}$.

The impact of the 
free-streaming of the WDM particles on the matter power spectrum  can be quantified through the transfer function $T(k)$ \citep{bode2001halo,markovic2011constraining}:
\begin{eqnarray}
T (k) = \left[ \frac{P_{\rm WDM} (k) } {P_{\rm CDM} (k) } \right]^{1/2} = \left[ 1 + (\alpha k)^{2 \mu} \right]^{-5/\mu} ,
\label{eq:eqpk}
\end{eqnarray}
Here $P_{\rm WDM} (k)$ and $P_{\rm CDM} (k)$ give
the linear matter power spectrum for the WDM and CDM model, respectively and 
$\alpha$, $\mu$ are the parameters that are used to model the transfer function $T(k)$ for the WDM model. We use the following fit for the parameters of
the transfer function$\mu$ and $\alpha$ \citep{viel2005constraining}: $\mu = 1.12$ and 
\begin{equation}
\alpha = 0.049 \, \left[ \frac{ m_{\rm wdm} } {{\rm keV} } \right]^{-1.11} \left[ \frac{ \Omega_{\rm wdm} } { 0.25 } \right]^{0.11} \left[ \frac{h} {0.7} \right]^{1.22} h^{-1} \, {\rm Mpc}
\label{eq:eqalpha}
\end{equation}
It follows from Eqs.~(\ref{eq:eqpk}) and~(\ref{eq:eqalpha}) that
as the $m_{\rm wdm}$ is decreased,   the  matter power is erased at progressively larger scales. In this work we consider WDM particles in the mass range:  $0.1 \, {\rm keV} <  m_{\rm wdm} < 4.6 \, {\rm keV}$.

Another well-studied alternative dark matter model arises from  ultra light axion (ULA) fields in the context of string axiverse \citep{Arvanitaki:2009fg,Marsh:2015xka,Marsh:2010wq,Hu:2000ke,Amendola:2005ad}.  The mass of ULA particles
lie in the range $m_a \simeq 10^{-33}\hbox{--}10^{-20} \, \rm eV$.

The ULA field  obtains its initial condition after the spontaneous symmetry breaking in the early universe and behaves like a coherent scalar field. At an early time when the expansion rate $H \gg m_a$, the ULA  behaves like a  cosmological constant. At redshifts when $m_a < H$, the field rolls off and starts oscillating coherently around the nearest minima of the periodic potential. During this
period, the average energy density of the field falls as $1/a^3$ and therefore
the background density acquires the characteristics of a CDM particle.

The adiabatic perturbations in the scalar field have a scale-dependent 
effective sound speed. At late times and sub-horizon scales, the effective sound speed is:
\begin{equation}
c_{\rm s}^2 = \frac{ \frac{k^2}{4 m_a^2 a^2} } { 1 + \frac{k^2}{4 m_a^2 a^2} }
\label{eq:eqcs}
\end{equation}
At scales  $k \gg m_a a$, the sound speed approaches the speed of
light which prevents clustering at these scales. At late times, the sound
speed approaches zero which causes the ULA density perturbation  to behave as
the cold dark matter perturbation. 

The  scale below  which the matter perturbations are suppressed can be approximated as  \citep{Marsh:2015xka}:
\begin{equation}
  k_m \simeq \Bigg(\frac{m}{10^{-33}eV}\Bigg)^{1/3} \Bigg(\frac{100\, \rm{km s^{-1}}}{c}\Bigg)h\, \rm{Mpc^{-1}}.
\end{equation}
A smaller $m_a$  yields  a larger  scale of  matter suppression. The cosmologically relevant mass  range  of ULA  is $10^{-25}\hbox{--}10^{-20} \, \rm eV$, which we consider here.

We  compute the ULA matter power spectra using the code \textit{AxionCAMB} which is the modified version of the
publicly-available code CAMB\footnote{available at https://github.com/dgrin1/axionCAMB}.

\begin{figure}[h]
\begin{center}
\vskip.2cm
\psfrag{k1}[c][c][1.0][0]{$k \, [h^{-1} {\rm Mpc}]^{-1}$} 
\psfrag{k2}[c][c][1.0][0]{$P(k) \, [h^{-1} {\rm Mpc}]^3$} 
\psfrag{CDM}[c][c][0.65][0]{$\Lambda$CDM \quad \quad \quad}
\psfrag{mwdm1}[c][c][0.65][0]{$m_{\rm wdm} = 1.0 \, {\rm keV}$ \quad \quad \quad }
\psfrag{mwdm1.5}[c][c][0.65][0]{$m_{\rm wdm} = 1.5 \, {\rm keV}$ \quad \quad \quad}
\psfrag{ma51e-23}[c][c][0.65][0]{$m_{\rm a} = 5 \times 10^{-23} \, {\rm eV}$ \quad \quad \quad \quad}
\psfrag{ma1e-23}[c][c][0.65][0]{$m_{\rm a} = 10^{-23} \, {\rm eV}$ \quad \quad \quad}
\centerline{\includegraphics[scale =.95]{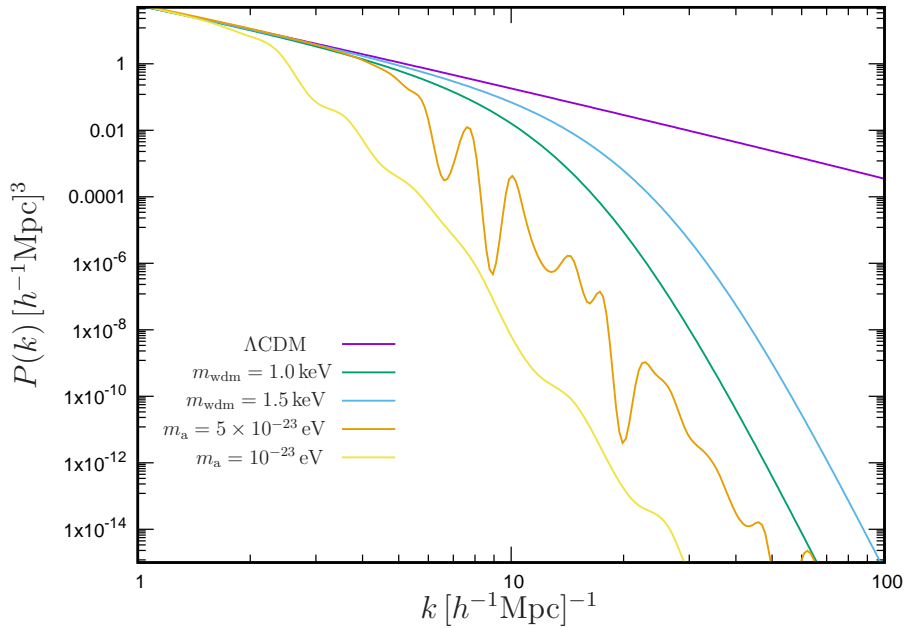}}
\caption{The figure displays the small-scale  normalized matter power spectrum for the  three  models  considered in this paper.  The choice of parameters for the different models is
listed in the inset.}
\label{fig:figpk}
\end{center}
\end{figure}

In Fig~\ref{fig:figpk}, we show the small-scale  matter power spectra for many WDM and ULA models  which are of interest to us in this paper. To motivate our discussion in the later sections, we note that for the  ULA models corresponding to:  $m_a = \{10^{-20}, 10^{-21}, 10^{-22}, 5\times 10^{-23}, 10^{-23} \} \, \rm eV$,
the matter power is 10\% lower than the $\Lambda$CDM model for $k \simeq \{30, 10, 4,3, 1.5\} \, \rm h \, Mpc^{-1}$, respectively. The corresponding numbers for the WDM models are:
  $m_{\rm wdm} = \{0.3, 1, 1.5, 2, 3 \} \, \rm keV$
 and $k \simeq \{0.3, 2, 3,4, 9\} \, \rm h \, Mpc^{-1}$, respectively.

\section{Simulating line-of-sight HI density fluctuations:  Lyman-$\alpha$ clouds} \label{sec:simudet}

Hydrodynamical simulations show that the Lyman-$\alpha$ clouds are mildly non-linear regions in the IGM  with density contrast  $\delta \leq 10$  at high redshifts. This allows one to capture the ionization, thermal, and dynamical state  of Lyman-$\alpha$  clouds
using semi-analytic  models.

In this paper, we have followed the semi-analytic approach given in
\citep{bidavidson1997}. We briefly outline different steps in this process. The first step is
the computation of  the three-dimensional baryonic matter power spectrum $P^{(3)}_{\rm b} (k, z)$ from the three-dimensional matter power spectrum $P^{(3)}_{\rm m} (k, z)$:  
\begin{equation}
P^{(3)}_{\rm b} (k, z) = \frac{P^{(3)}_{\rm m} (k, z)}{ (1+\lambda_b^2k^2)^2 },
\label{eq:pk3d}
\end{equation}
Here $\lambda_b \equiv k_J^{-1}$ is the thermal Jeans scale:
\begin{equation}
 \lambda_b  = \frac {1}{H_0} \left[ \frac{ 2 \gamma k_{\rm B} T_{\rm m} (z) }{ 3 \mu m_{p} \Omega_{\rm m} } \right]^{1/2} \times 
(1+z)^{-1/2}
\label{eq:jeans}
\end{equation}
All the parameters that appears in Eq.~(\ref{eq:jeans})  have their usual meanings with  $\mu= 0.6$. For the parameters we use,  $k_J \simeq  8.5 \, h \, {\rm Mpc}^{-1}$ at $z = 3$\footnote{ The thermal Jeans' scale correspond to
  Jeans' mass $M_J \simeq 5\times 10^8 \, \rm M_\odot$ at $z\simeq 3$. The Jeans' mass in the unheated IGM at
  $z\simeq 20$ is $M_J \simeq 10^5 \,\rm M_\odot$. After the re-heating and the reionization of the IGM in the redshift range $7 <z < 20$, the Jeans' mass increases and remains large in the post-reionization universe. The net impact of these processes is to prevent of growth of perturbations at sub-Jeans' scale for $z \lesssim 20$. While this would result in the  suppression of the density contrast at these scales as compared to the  larger scales, the suppression might  not be as large as given in Eq.~(\ref{eq:pk3d}), which corresponds to a more conservative choice. This is partly justified as the reheating and reionization history of the universe during the era of cosmic dawn and the reionization is not yet well understood. \label{fn:jeans}}

The relevant one-dimensional baryonic power spectra can be computed from the
three-dimensional  baryonic  power spectrum (Eq.~\ref{eq:pk3d}):
\begin{eqnarray}
P^{(1)}_{\rm b} (k_1, z) = \frac{1}{2 \pi} \int_{|k_1|}^{\infty} dk^{'} k^{'} P^{(3)}_{\rm b} (k^{'}, z) \nonumber \\
P^{(1)}_{\rm v} (k_1, z) = \dot{a}^2 k^2 \frac{1}{2 \pi} \int_{|k_1|}^{\infty} \frac{ dk^{'} }{ k^{'3} } P^{(3)}_{\rm b} (k^{'}, z) \nonumber \\
P^{(1)}_{\rm bv} (k_1, z) = i \dot{a} k \frac{1}{2 \pi} \int_{|k_1|}^{\infty} \frac{dk^{'}}{k^{'}} P^{(3)}_{\rm b} (k^{'}, z)
\label{eq:powspec_a}
\end{eqnarray}
Here $k_1$  denotes the Fourier mode along the line of sight. $P^{(1)}_{\rm b} (k, z)$, $P^{(1)}_{\rm v} (k, z)$ and $P^{(1)}_{\rm bv} (k, z)$ denote  the line-of-sight  density, velocity, and  the cross power spectra of  the density and the velocity fields,  respectively.

Given the one-dimensional power spectra (Eq.~(\ref{eq:powspec_a}), the correlated density and velocity fields are simulated  by applying the Gram-Schimdt procedure on  two
independent Gaussian random fields (for details see e.g. \cite{2013ApJ...762...15P} and references therein).

This process yields the  correlated density and velocity fields in the Fourier space.  we then apply inverse Fourier transform on these fields to obtain  the corresponding  real-space density and velocity fields, $\delta_b(x,z)$ and $v(x,z)$,  respectively. The main aim of this paper is to study the redshift evolution of
observables derived from Lyman-$\alpha$ data. Therefore,  we  simulate the density and the velocity fields for 40 different redshift bins each  of  width $\Delta z = 0.1$ over the redshift range $2 \leq z \leq 6$. In each of these bins, there are $2^{14}$ points that resolve  the Jeans scale $\lambda_b$  by at least a factor of 4. The simulated fields are then smoothed to the instrumental
resolution (discussed in detail later). 

In the semi-analytic approach, the non-linearity of the density perturbations in the IGM is incorporated by assuming the density field to follow  the lognormal distribution  \citep{bidavidson1997}. This allows us to relate  the
simulated baryonic  density contrast, $\delta_{\rm b} (x, z)$ to the baryon number density $n_{\rm b} (x, z)$ as, 
\begin{equation}
n_{\rm b} (x, z) = A e^{ \delta_{\rm b}(x, z) }
\label{eq:nb} 
\end{equation}
$A$ is a  normalization constant that can be determined by averaging the baryon number density: 
\begin{equation}
\langle n_{\rm b} (x, z) \rangle \equiv n_0(x, z) = A \langle e^{ \delta_{\rm b}(x, z) } \rangle
\label{eq:norm} 
\end{equation}
Since the density perturbation $\delta_{\rm b} (x,z)$ is assumed to be Gaussian, we can write $\langle e^{ \delta_{\rm b} (x, z) } \rangle$ as,
\begin{equation}
\langle e^{ \delta_{\rm b} (x, z) } \rangle = e^{ \langle \delta^2_{\rm b} (x, z) \rangle / 2 }
\label{eq:exp}
\end{equation}
Using Eqs.~(\ref{eq:norm}) and~(\ref{eq:exp})  in Eq.~(\ref{eq:nb}), the baryon number density $n_{\rm b} (x, z)$ can be expressed as:
\begin{equation}
n_{\rm b} (x, z) = n_0 (z) e^{ ( \delta_{\rm b} (x, z) - \langle \delta^2_{\rm b} (x, z) \rangle / 2 ) } 
\end{equation}
where $n_0 (z)$ gives the background number density of the baryons at a redshift $z$:
\begin{equation}
n_0 (z) = \frac{\Omega_{\rm b} \rho_c}{\mu_b m_p } (1+z)^3 
\end{equation}
Here $\mu_b = 1.2$. 

\section{Neutral hydrogen and  the Lyman-$\alpha$ optical depth }
\label{sec:obsqua}

For comparison with Lyman-$\alpha$ data, we need the fluctuating component
of the neutral hydrogen field along the line of sight, $n_{\rm HI}(x,z)$. The
relation between the baryonic field (Eq.~(\ref{eq:norm})) and the neutral hydrogen field can be computed by assuming 
ionization equilibrium in the IGM and the Lyman-$\alpha$ clouds, which are optically thin to the ionizing radiation. This gives us:
\begin{equation}
  n_{\rm HI}(x, z) \simeq  \frac{\alpha[T(x,z)] n_{\rm b}(x,z) } {\Gamma_{ci} (x,z) + J(z)/[\mu_e n_{\rm b}(x,z)] }
  \label{eq:neu_hy}
\end{equation}
In writing Eq.~(\ref{eq:neu_hy}), it has been implicitly assumed that the
gas is highly ionized, as is expected for the  physical parameters of the IGM.
The temperature field,  $T(x,z) = T_0(z) [ n_{\rm B} (x,z) / n_0(z) ]^{\gamma - 1}$, where $T_0$ gives the mean IGM temperature and $\gamma$ is the polytropic index of  the gas. The  variation of $\gamma$  provides information about the  dynamical state of the Lyman-$\alpha$ clouds.  $\alpha[T(x,z)]$ is the recombination coefficient and its value is determined by the temperature field at a given  location. $\Gamma_{\rm ci}(x,z)$ is the coefficient of collisional ionization (in $\rm cm^3 \, sec^{-1}$) and $J(z)$ 
denotes the rate of photoionization (in $\rm sec^{-1}$). For a fully ionized gas,
$\mu_e = 1.07$ (see Eq.~(25) in \cite{choudhury2001semianalytic}). To model Lyman-$\alpha$ clouds, 
the free  parameters have values in the ranges: 
$7000 \leq T_0 \leq 15000$ K and $1.3 \leq \gamma \leq 1.6$ \citep{hui1997}.

To compare the simulated data set against the observed flux
decrement in the Lyman-$\alpha$ clouds,
we need the optical depth $\tau(\nu)$ of the clouds for Lyman-$\alpha$ scattering as a function of frequency, which   is given by:
\begin{equation}
  \tau(\nu) = \int n_{\rm HI}(x,z) \sigma_a\left( \frac{\nu}{a} \right) dl.
  \label{eq:opdep}
\end{equation}
Here  $n_{\rm HI}(x,z)$ gives the number density of the HI and  $\nu$ is the observed frequency.  The absorption cross section $\sigma_a$ is given by:
\begin{equation}
\sigma_a = \frac{I_a}{ b \sqrt{\pi} } V \left( \alpha, \frac{\nu - \nu_a}{b \nu_a} + \frac{v}{b} \right),
\end{equation}
where the parameter $b = (2 k_{\rm B} T / m_p)^{1/2}$ gives the velocity dispersion of the HI atoms, $v(x)$ is   HI velocity field, $\alpha = 2 \pi e^2 \nu_a / 3 m_e c^3 b = 4.8548 \times 10^{-8} / b$, $I_a = 4.45 \times 10^{-18}$ cm$^{-2}$, and  $V$ denotes the Voigt function.

For a temperature $T \simeq 10^4$ K, these effects combine to yield us the optical depth $\tau$ as a function of redshift, which can be approximated as (e.g. \citep{croft1998}):
\begin{equation}
\tau(z) = A (n_{\rm B} / n_0)^{ 2 - 0.7(\gamma-1) }
\label{eq:tau}
\end{equation}
where 
\begin{equation}
  A = 0.946 \left( \frac{1+z}{4} \right)^6 \, \left( \frac{\Omega_b h^2}{0.0125} \right)^2 \, \left( \frac{T_0}{10^4 {\rm K}} \right)^{-0.7} \, \left( \frac{J}{10^{12}{\rm s}^{-1}} \right)^{-1} \, \left( \frac{H(z)}{H_0} \right)^{-1}
  \label{eq:anormta}
\end{equation}

For the purpose of  comparison with the data, we  compute  the Lyman-$\alpha$ effective optical depth $\tau_{\rm eff} (z)$ as,
\begin{equation}
  \tau_{\rm eff} (z) = -{\rm log}[ \langle {\rm exp}(-\tau )\rangle ],
  \label{eq:opdepens}
\end{equation}
which is the observable that quantifies the decrease in the observed flux ($ F \propto e^{-\tau} $) as a function of redshift. The closed angular bracket $\langle ..\rangle$ in the above equation refers to the average of ${\rm exp}(-\tau)$ over all possible realizations of the optical depth  $\tau \equiv \tau(z)$ for a given redshift bin.

\subsection{ Data for $\tau_{\rm eff}$ }
The aim of this paper is to constrain certain dark matter models in which the
small scale power ($\leq \hbox{a few}\, {\rm Mpc}$) is suppressed in comparison with   the usual $\Lambda$CDM model.  This motivates us  to analyze  high-resolution observations of
the Lyman-$\alpha$ forest. We therefore consider the data compiled by \cite{faucher2008}. This data 
measures  of the Lyman-$\alpha$ optical depth $\tau_{\rm eff}$ over
a redshift range $2 \leq z \leq 4.2$ at an interval $\Delta z = 0.2$
using a sample of 86 high-resolution, high S/N (Signal-to-Noise) quasar spectra. The
data set consists of  16 (44) quasar spectra observed using  HIRES (ESI) spectrographs with the  Keck telescope, while the remaining  26 quasar spectra were obtained from   with the  MIKE instrument  on
Magellan. The HIRES and ESI spectrographs
yield velocity resolutions  $\Delta v \simeq 6\hbox{--}8 \, {\rm km/s}$ and  $\Delta v \simeq
33\hbox{--}44  \, {\rm km/s}$, respectively, and the MIKE spectrograph
operates at $\Delta v \simeq 11\hbox{--}14 \, {\rm km/s}$. Most of the  quasar
spectra obtained using HIRES and ESI spectrographs are used to study
the Damped Lyman Alpha (DLA) systems
\citep{prochaska1999,prochaska2007} and have ${\rm S/N} \geq 15 \,
 {\rm pixel}^{-1}$, whereas the spectra obtained using MIKE
  spectrograph were analyzed for the Super Lyman Limit (SLL)
  systems given in \cite{meara2007} and has ${\rm S/N} \geq 10 \,
  {\rm pixel}^{-1}$.

For an expanding universe, the velocity resolution width $\Delta v$ corresponds to a line-of-sight
comoving length: $\Delta r = \Delta v (1+z) / H(z)$. This 
gives $\Delta r \simeq 0.01 \, {\rm Mpc}$ for  $\Delta v = 1 \, {\rm km/s}$ at $z = 3$.  This allows us to express  the observed velocity resolution in
terms of the largest  line-of-sight Fourier mode that the data  can probe (Eq.~(\ref{eq:powspec_a})).  One can readily show that the lowest resolution of
the data we use ($\Delta v \simeq 40 \, \rm km/s$) is comparable to
the Jeans' scale (Eq.~\ref{eq:jeans}). In our theoretical model, the baryonic power spectrum is sharply cut off at the Jeans' scale (Eq.~(\ref{eq:pk3d})), which means that even though the data resolves the Jeans' scale  we do not expect the data to contain information on scales $k > k_J$. In our analysis, we first
obtain the optical depth for individual clouds (Eq.~(\ref{eq:opdep})), which is followed by
the  computation of the transmitted flux  $\propto \exp(-\tau_i(\nu))$  as a function of frequency.  The simulated  transmitted fluxes   are  then smoothed
with a velocity resolution $\Delta v = 7.5 \, {\rm km/s}$ for comparison with data.   
We note that   our results are insensitive to the choice of the velocity
  resolution if the length scale corresponding to the velocity resolution
  is smaller or comparable to the thermal Jeans' scale (Eq.~\ref{eq:jeans}), which is the smallest scale we can probe in our study. 

More Recently, \cite{becker2013} have estimated  $\tau_{\rm eff}$ over a larger redshift range $2 \leq z \leq 5$ using a sample
of 6065 moderate-resolution quasar spectra drawn from SDSS DR7 \cite{schneider2010}. However, the velocity resolutions  of
the quasar spectra are in the range  $\Delta v \simeq 143 \hbox{--} 167 \, {\rm km/s}$, which 
correspond to length scales considerably larger than the Jeans' scale and therefore are not  suitable for our analysis\footnote{ We note the mean flux decrement is independent of  spectral resolution and therefore in principle our proposed method should work with low-resolution data sets also. However, continuum subtraction does depend on spectral resolution and it impacts the mean flux decrement. Hence,  we prefer to work with the  highest spectral resolution data set. }

\section{Results} \label{sec:resu}

Our analysis is based on comparing  the measured effective optical depth
against the simulated optical depth as a function of redshift.
We compute the effective optical depth  $\tau_{\rm eff}(z)$
from the simulated line-of-sight HI field  for three different cosmological models --- 
$\Lambda$CDM, WDM and ULA models. The dark matter model is constrained from
comparing the slope of the redshift dependence of the effective optical depth
for different models. 

 We simulate  the Lyman-$\alpha$ effective
optical depth $\tau_{\rm eff} (z)$  over the redshift range
$2 \leq z \leq 6$ at a redshift interval of $\Delta z = 0.1$
using the method detailed in the last section.  Throughout this section, the following
parameters, unless otherwise specified,  are used for computing  $\tau_{\rm eff} (z) $:  $ J = 1.5 \times 10^{-12} \, {\rm s}^{-1}$, $T_0 = 1.3 \times 10^{4} \, {\rm K}$, $\gamma = 1.4$, and 
$\Delta v = 7.5 \, {\rm km/s}$.

The Lyman-$\alpha$ effective optical depth
$\tau_{\rm eff} (z)$ at a given redshift $z$ can be calculated by replacing the ensemble average defined in Eq.~(\ref{eq:opdepens}) by an average over optical depths of individual clouds in the simulated HI field: 
$\tau_{\rm eff} (z) = -{\rm log} [ \sum \limits_{i} {\rm exp}
  (-\tau_i) / N]$ where $N$ and $\tau_i$ denote the number of the
Lyman-$\alpha$ clouds and the optical depth of the $i$-th
Lyman-$\alpha$ cloud  at that redshift, respectively.  The HI column  density
of a Lyman-$\alpha$ cloud at a given redshift depends on
both the baryonic  density perturbation $\delta$ and the background
baryon  density at that redshift.  In a matter-dominated universe, a reasonable
assumption for the standard $\Lambda$CDM model at $z > 2$, the matter density
contrast,   $\delta \propto (1+z)^{-1}$ while the background density 
 $n_{\rm HI} \propto (1+z)^3$. This gives:
$\tau_i \propto n_{\rm HI}^2 H^{-1}(z) (1+ \delta)^{\beta}$ where $\beta = 2 - 0.7(\gamma - 1)$. Using
$\delta \gg 1$, we get, $\tau_i \propto (1+z)^{4.5 - \beta}$. Here 
 $H(z) \propto \Omega_m^{0.5}(1+z)^{1.5}$  is the Hubble parameter. The optical depth of an  individual
Lyman-$\alpha$ cloud,  $\tau_i$, evolves   sharply with the redshift. A part
of this change  arises from the evolution of background quantities (Hubble's parameter and background density) but an additional change  comes from the density contrast and equation of state $\gamma$  of the clouds. 

At any given redshift, the range of  optical depths of  clouds
is governed by the log-normal density distribution (Eq.~(\ref{eq:nb})).
To understand the redshift evolution of this distribution, we  note
that the line-center cross-section for Lyman-$\alpha$ scattering for a non-expanding region of temperature $T\simeq 10^4 \, \rm K$ is $\simeq 5 \times 10^{-14} \, \rm cm^2$. As the column densities of most  clouds lie  in the range $10^{13}\hbox{--}10^{15} \, \rm cm^2$, many clouds are optically thick at any redshift
and a larger fraction of clouds is  optically thick at higher redshifts.

To further analyze the implication of this dynamics, we compare the redshift evolution of
$\tau_{\rm avg}(z) \equiv \sum \limits_i \tau_i/N$ with the observationally determined effective optical depth,  $\tau_{\rm eff} (z)$. If optical depths of all the  Lyman-$\alpha$ clouds  are small, $\tau_i \ll 1$, then $\tau_{\rm avg} \simeq  \tau_{\rm eff}$  at all redshifts. In the other limit, if all the clouds are optically thick  $\tau_i > 1$,
the redshift evolution of $\tau_{\rm avg}$ is the same as in the first case, but the $\tau_{\rm eff}$ doesn't change with the  redshift. In the intermediate situation which is suitable for
the allowed cosmological models, a fraction of clouds are optically thick
at any redshift and this fraction increases at larger redshifts. This implies
that the redshift evolution of $\tau_{\rm eff}$ is always flatter than the
redshift dependence of $\tau_{\rm avg}$. 

\begin{figure}[h]
\begin{center}
\vskip.2cm \psfrag{k2}[c][c][1.0][0]{redshift $z$}
\psfrag{k1}[c][c][1.0][0]{$\tau (z)$} 
\psfrag{tauavg}[c][c][0.65][0]{$\tau_{\rm avg} (z)$} 
\psfrag{taueff}[c][c][0.65][0]{$\tau_{\rm eff} (z)$}
\centerline{\includegraphics[scale =.95]{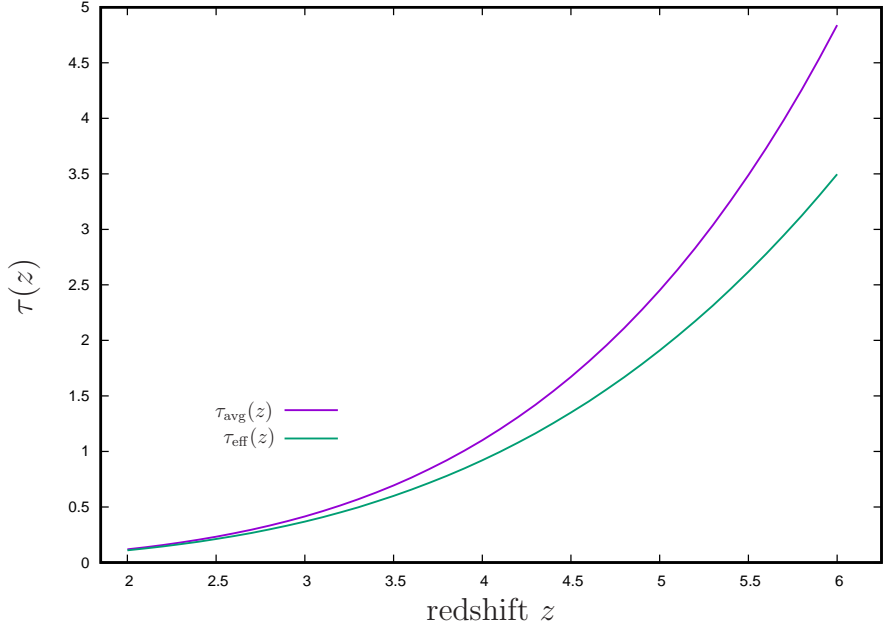}}
\caption{The figure shows the redshift evolution of the effective and the average optical depths, $\tau_{\rm eff} (z)$ and $\tau_{\rm avg}(z)$,  for the $\Lambda$CDM model for the following  parameters: $ J = 1.5 \times 10^{-12} \, {\rm s}^{-1}$, $T_0 = 1.3 \times 10^{4} \, {\rm K}$, $\gamma = 1.4$ and 
$\Delta v = 7.5 \, {\rm km/s}$. }
\label{fig:fig1}
\end{center}
\end{figure}

In figure~\ref{fig:fig1}, we show the redshift dependence of  Lyman-$\alpha$ effective optical depth
$\tau_{\rm eff}(z)$ and the average Lyman-$\alpha$ optical depth
$\tau_{\rm avg}(z)$ for the $\Lambda$CDM model. We notice  that     
$\tau_{\rm eff}(z) \simeq \tau_{\rm avg} (z)$ at $z\simeq 2$ but $\tau_{\rm ave} (z)$ diverges from  $\tau_{\rm eff}(z)$ at high redshifts. As anticipated, the slope of $\tau_{\rm eff}(z)$ is smaller than that of  $\tau_{\rm avg} (z)$.

\subsection{Effective optical depth and dark matter models}

The main aim of this paper is to constrain dark matter models using
the redshift evolution of $\tau_{\rm eff}$. The alternative dark matter
models we consider are described in detail in section~\ref{sec:dmmod}. The discussion in the foregoing
provides motivation of this aim and we elaborate it further here.

Both the alternative models we consider yield a suppression of matter power
at small scales as compared to the usual $\Lambda$CDM model (Figure~\ref{fig:figpk}). The scales at which the power
is suppressed is determined by the mass of particles in both cases.
As discussed above, the optical depth of an individual cloud depends on
both the background density and the density contrast, $\delta$. While the background
density is left unchanged in  alternative dark matter models we consider, the density contrast decreases  owing to the decrement in  matter power at small scales. 

\begin{figure}[h]
\begin{center}
\vskip.2cm \psfrag{k3}[c][c][0.8][0]{redshift $z$}
\psfrag{k1}[c][c][0.8][0]{$\tau_{\rm avg}(z)$}
\psfrag{k2}[c][c][0.8][0]{$\tau_{\rm eff}(z)$} 
\psfrag{CDM model}[c][c][0.5][0]{$\Lambda$CDM \quad \quad  } 
\psfrag{WDM model - mwdm = 1.0 keV}[c][c][0.5][0]{\quad \quad \quad $m_{\rm wdm}  = 1.0 \, {\rm keV}$  }
\psfrag{Axion model - ma = 1e-23 keV}[c][c][0.5][0]{\quad \quad \quad \quad $m_{\rm a} = 10^{-23} \, {\rm eV}$  }
\centerline{{\includegraphics[scale =0.60]{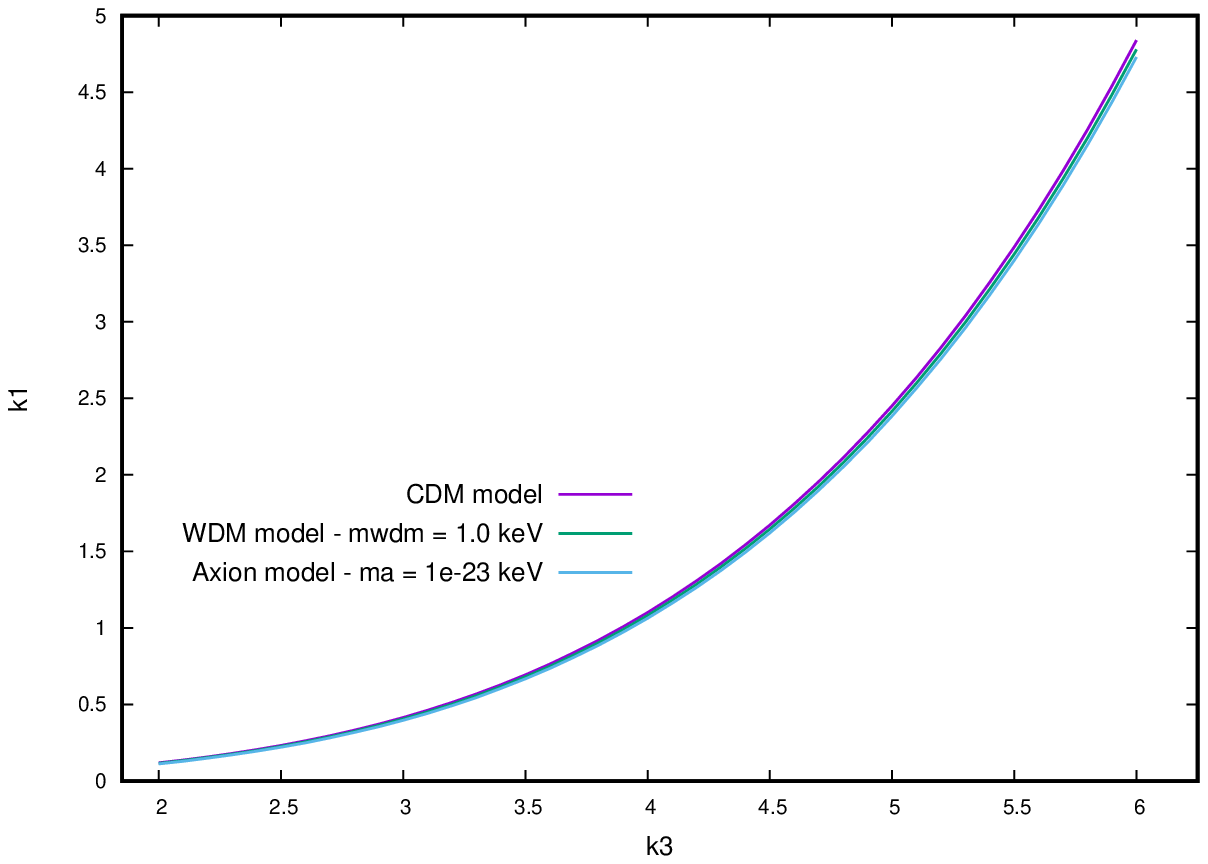}}
  {\includegraphics[scale=0.60]{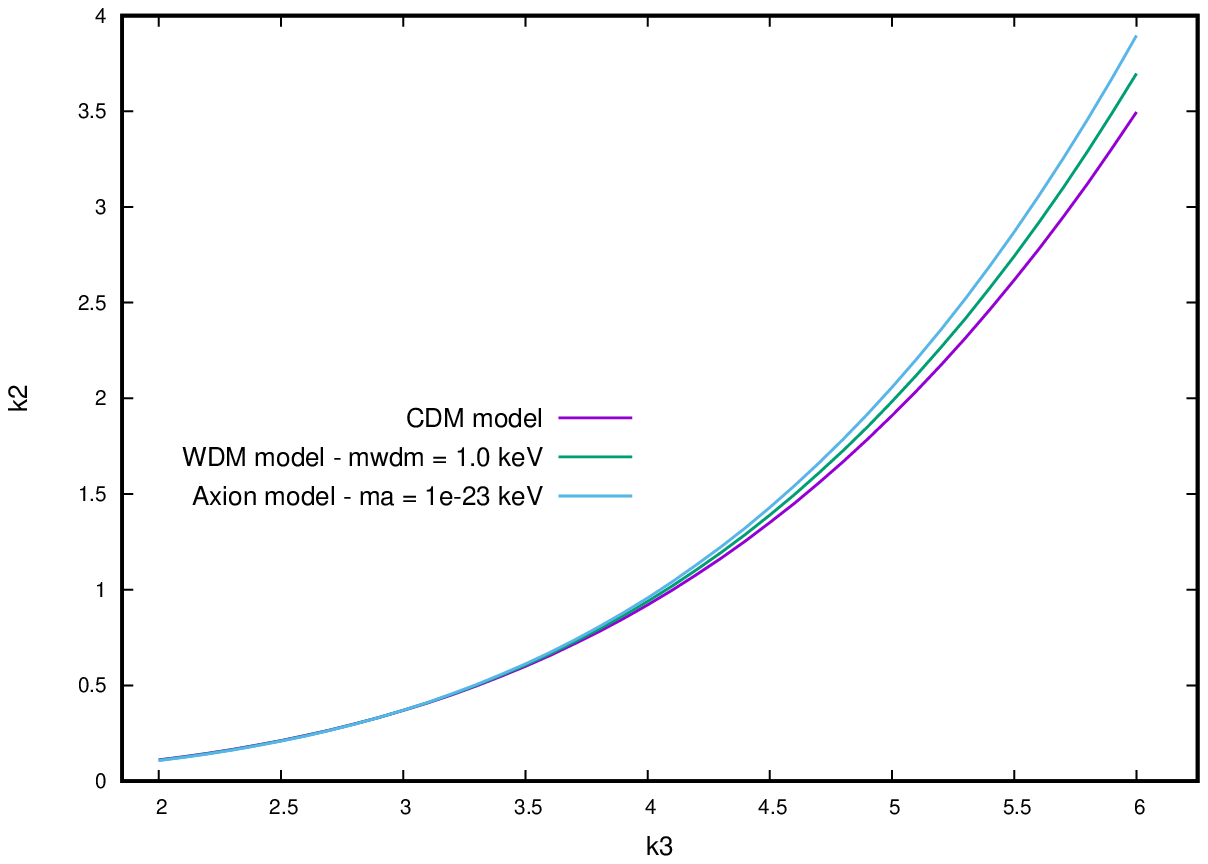}}}
\caption{The figure displays the redshift evolution of the
  average and effective optical depths, $\tau_{\rm avg} (z)$ and $\tau_{\rm eff} (z)$, for $\Lambda$CDM, WDM, and ULA models. The other modelling parameters are the same as in Figure~\ref{fig:fig1}.}
\label{fig:fig3}
\end{center}
\end{figure}

We next study the evolution
of  $\tau_{\rm eff}(z)$ and  $\tau_{\rm avg}(z)$  in these models relative to the $\Lambda$CDM model, which is  shown in Figure~\ref{fig:fig1}.  In Figure~\ref{fig:fig3}  the
relevant plots are displayed. As expected $\tau_{\rm avg}(z)$ is smaller in alternative
dark matter models  because of the decrement of density
contrast in these models. The difference between the average optical depth 
$\tau_{\rm avg} (z)$ for these models and that for the $\Lambda$CDM
model is  seen to  be negative in  the redshift
range $2\leq z \leq 6$.

This also allows us to understand
the difference of the average optical depth between  the 
two alternative dark matter models. The scales at which the
suppression of the  matter power occurs in the  WDM model  for $m_{\rm wdm} = 0.3 \, {\rm keV}$ are  larger  as compared to the ULA model for  $m_{\rm a} = 10^{-23} \, {\rm eV}$. The matter power
gets suppressed roughly by a factor $\simeq 10$ in the range 
$k \simeq 0.3 \hbox{--} 0.5 \, h \, {\rm Mpc}^{-1}$  for $m_{\rm wdm} = 0.3 \, {\rm keV}$, whereas, for $m_{\rm a} = 10^{-23} \, {\rm eV}$, the matter
power decreases by a factor $\simeq 3$ for  the range $k \simeq 0.3\hbox{--}1.0 \, h \, {\rm Mpc}^{-1}$.  This explains the smaller values of  $\tau_{\rm avg}(z)$ for the WDM model for  $m_{\rm wdm} = 0.3 \, {\rm keV}$ as compared to
the ULA models for $m_{\rm a} = 10^{-23} \, {\rm eV}$ over the entire redshift range. 

Figure~\ref{fig:fig3} also shows the effective optical depth for different
models. However, unlike the monotonic behavior of average optical depth,  we notice a transition 
at $z\simeq 3$\footnote{The redshift at which the transition occurs depends
  on the choice of dark matter model parameters.  For instance, the transition redshift shifts to smaller redshifts as $m_{\rm wdm}$ is increased.}. The discussion in the foregoing allows us to understand this
transition. At small redshifts, the $\Lambda$CDM model has higher effective
optical depth for the following reasons: (a) the optical depth for an individual cloud, $\tau_i < 1$,  for most clouds for all the models and (b)  the optical depth of individual clouds
$\tau_i$ is higher for the $\Lambda$CDM model owing to a larger density contrast. So the effective optical depth converges to the average optical depth at small
redshifts (Figure~(\ref{fig:fig1})), which is higher for the $\Lambda$CDM models
as compared to the alternative dark matter models. This also means that at higher redshifts a larger number of clouds become optically thick for the $\Lambda$CDM which causes the effective optical depth to partially  saturate and consequently the
redshift evolution of the effective optical depth has a smaller slope. On the other hand,  for
the alternative dark matter models,  the effective optical depth
tracks the average optical depth up to a much higher redshifts and therefore has
a sharper slope in the redshift space. This explains the transition from
a positive difference to a negative difference in the effective optical depth between the $\Lambda$CDM and the alternative dark matter models. This also explains the evolution of the  difference 
between the two alternative dark matter models.

To illustrate this aspect further, we 
show the relative difference of the average and effective optical depths between the $\Lambda$CDM model and 
the alternative dark matter models in Figure~\ref{fig:fig4}. 
We define the relative difference $\Delta \tau (z)$ as, 
$\Delta \tau (z) = [ \tau^{\rm model} (z) / \tau^{\rm CDM} (z) - 1 ]\times 100 \%$ for a given model where $\tau^{\rm CDM}$ and
$\tau^{\rm model}$  refer  to corresponding quantities  for the $\Lambda$CDM
model and one of the alternative dark matter models considered in our analysis. As expected , the relative
differences $\Delta \tau_{\rm avg} (z)$ are negative throughout the
redshift range $2 \leq z \leq 6$ (see the left panel of Figure~\ref{fig:fig4}).  
For the ULA model with 
$m_{\rm a} = 10^{-23} \, {\rm eV}$, the relative difference  varies in the range  $\Delta \tau_{\rm avg}(z) \simeq 3 \hbox{--} 6 \%$ over the redshift range $2 \leq z \leq 6$. For the WDM model with $m_{\rm wdm} = 0.3 \, {\rm keV}$, 
the difference increase by factors $\sim 1.3 \hbox{--} 1.6$ over the redshift range.

We further see that the relative differences $\Delta \tau_{\rm eff}$ are negative at
redshifts $z < 3$ and positive at redshifts $z > 3$ (see the right panel of Figure~\ref{fig:fig4}).  
For $m_{\rm a} = 10^{-23} \, {\rm eV}$, the relative
difference varies from   $\Delta \tau_{\rm eff} (z) \sim -4 \%$ to $\Delta \tau_{\rm eff}(z) \sim 12 \%$ over
the redshift range $2 \leq z \leq 6$.  The WDM models show similar cross-over
from negative to positive difference as is seen in the Figure~\ref{fig:fig4} (right panel). 
We note that for the WDM model with $m_{\rm wdm} = 0.3 \, {\rm keV}$, the difference 
increases by a factor $\sim 1.5$ over the redshift range.

\subsection{Comparison with data}

In Figure~\ref{fig:fig5} we display the redshift evolution of $\tau_{\rm eff}(z)$ for a few dark matter models along with the high-resolution Lyman-$\alpha$ data (\cite{faucher2008}).

For a detailed statistical analysis, we carry out a likelihood analysis
that compares the simulated  effective optical depths
with the data. 

In addition to the masses: $m_{\rm wdm}$
and $m_{\rm a}$, we  consider the following four parameters related to the ionization, thermal and dynamical state of Lyman-$\alpha$ clouds in our analysis:
$J(z)$, $\alpha$, $T_0$ and $\gamma$.

The ionization
and thermal history in the redshift range $2 \lesssim z \lesssim 5$ is known
from the observation of Lyman-$\alpha$ clouds 
(for a comprehensive review see e.g. \cite{mcquinn2016evolution} and references therein). The temperature of the IGM 
$T_0 \simeq 10^4 \, \rm K$  with no discernible redshift evolution  for $z > 3.5$. The reionization of singly-ionized helium raises the temperature
by approximately a factor of 1.5 for $z < 3.5$ \cite{plante2018helium}. This is consistent with the
fact that the temperature of optically-thin photoionized hydrogen gas
is nearly independent of  the magnitude of the ionizing flux with a weak dependence on the  spectral index 
\cite{draine2010physics,mcquinn2016evolution}. 
We therefore 
expect $T_0$ to be nearly independent at higher redshifts if the gas is
photoionized and the spectral index of the ionizing sources doesn't change
appreciably. The equation of state parameter $\gamma$ has been  extensively
studied using hydrodynamical simulations and its acceptable range has been
determined by comparison with Lyman-$\alpha$ forest data 
\cite{hui1997,lukic2015lyman}.
The Lyman-$\alpha$ forest data suggests that the ionizing intensity  $J(z)$ 
doesn't evolve in the redshift range $2 < z < 4.2$  \cite{faucher2008flat}. 
However, its redshift evolution at higher redshifts remains uncertain. 
We account for the redshift evolution of the ionizing intensity $J(z)$ by assuming:
$J(z) = J_0 (1+z)^{\alpha}$; here $J_0$ and $\alpha$ are constants.

Based on these considerations, we vary $J_0$, $\alpha$, $T_0$ and $\gamma$ over the following ranges (with a flat prior) in the likelihood analysis:  $0.7 \times 10^{-12} \leq J_0 \leq 4 \times 10^{-12}$, $-1.2 \leq \alpha \leq 1$, 
$0.7 \times 10^4 \leq T_0 \leq 2.3 \times  10^4$ and
$1.4 \leq \gamma \leq 2.2$. For the WDM and  ULA models, the masses $m_{\rm wdm}$ and $m_{\rm a}$ are 
varied over the ranges: $0.1 \, {\rm keV} \leq m_{\rm wdm} \leq 4.6 \, {\rm keV}$ 
and $10^{-24} \, {\rm eV} \leq m_{\rm a} \leq 5 \times 10^{-22} \, {\rm eV}$. 
  Before embarking on the details of likelihood analysis, we
  discern the structure of degeneracy of the proposed five parameter likelihood  function.
  Eqs.~(\ref{eq:tau})--(\ref{eq:opdepens}) show the dependence of effective
  optical depth on all the parameters. The alternative dark matter impacts the effective optical depth  through the baryon density
  $n_b \equiv n_0(1+\delta_B)$ in Eq.~(\ref{eq:tau}). For mildly non-linear regions,
  $\delta_B \simeq 10$, $n_b \simeq n_0 \delta_B$. For alternative
  dark matter models, $\delta_B$ is obtained by simulating the density
  and velocity fields from a different matter power
  spectrum. So if  a combination of the other modelling parameters/functions, needed to
  infer the density field from  the observed flux decrement 
  ($\gamma$, $T_0$, and $J(z)$),  can entirely compensate for the change in $\delta_B$ over the redshift range of interest, the proposed impact of the change in the cosmological model would be unobservable.  Eqs.~(\ref{eq:tau})--(\ref{eq:opdepens}) also allow us to compare the efficacy of our
  approach with the more direct method, which  is normally based
  on the simulation of the  three-dimensional density field using hydrodynamical simulations.  This allows one to draw the density field along  many  line-of-sights. Eqs.~(\ref{eq:tau})--(\ref{eq:opdepens})  yield the line-of-sight transmitted flux field as a function of frequency, which  is then used to
  compute transmitted flux correlation function. The  correlation
  function from the simulated data  can then  be compared to the Lyman-$\alpha$ data. As  Eqs.~(\ref{eq:tau})--(\ref{eq:opdepens}) are needed to compare the
  theoretical predictions with the data for this method also, it  also partly shares
  the degeneracy structure  of our proposed  method. 
  
  However, while our method is based on using just one quantity, average overdensity,   at a given
  redshift, the other  method computes  flux correlation function  at any redshift
  and therefore is  also sensitive to the spatial variation of the density field
  at any given redshift, while it can also use information from different redshifts.   This method, therefore,  retains  more information than our approach. And,  given the uncertainty in modelling parameters,  it is expected to
  yield better constraints on the dark matter masses.

It is difficult to quantify the nature of the degeneracy  analytically as   the average overdensity is computed from a simulation and $\tau_{\rm eff}$ is a highly non-linear function of modelling parameters.  Eq.~(\ref{eq:tau}) only partially captures the nature of this
  effect as the Lyman-$\alpha$ clouds that most impact the redshift evolution of
  $\tau_{\rm eff}$ are the ones with $\tau \simeq 1$ (Figure~\ref{fig:fig4}). However, Eq.~(\ref{eq:tau})
  strongly suggests that the only models that can be constrained using our
  method are the ones for which the change in overdensity is larger than obtained  by changing the  modelling parameters within the prior range. This also means that we expect the bound on dark matter mass to weaken as the prior range is expanded. To study 
  this further, we explored the dependence of the dark matter posterior probability  on the modelling parameters by expanding  the range of priors for each parameter. We found that  the dependence on the prior is negligible if the maximum of
  the posterior probability  of the relevant parameter lies in the prior range. For $\gamma$ the maximum of the  posterior probability is around 1.1 which is outside the original range ($1.4 \leq \gamma \leq 2.2$). 
We extended the prior range to encompass the maximum ($0.5 \leq \gamma \leq 2.2$)
 and find a discernible but small  weakening of the bound on dark matter masses. This also means
  that a better understanding of priors from complementary data might  allow better determination 
of dark matter masses using our method.

Our analysis yields likelihood function as a function of the parameters, $J_0$, $\alpha$, $T_0$, $\gamma$, and either of the two masses, $m_{\rm wdm}$ or $m_{\rm a}$. As our method of constraining the alternative
    dark matter models relies upon the
    redshift evolution of $\tau_{\rm eff}$, we first investigate  the possible
    degeneracy between the masses of alternative dark matter particles
    and $\alpha$, which gives the redshift evolution of the ionizing radiation.
    We marginalize over the parameters $J_0$, $T_0$, $\gamma$ to obtain the marginalized joint likelihood
of $\alpha$ and the masses: $m_{\rm wdm}$ and $m_{\rm a}$, for the WDM and the ULA models respectively. 
We find that the parameters: $\alpha$ and the masses: $m_{\rm wdm}$ and $m_{\rm a}$ are weakly anti-correlated. We marginalize the joint likelihood over the masses: $m_{\rm wdm}$ and $m_{\rm a}$ to get the marginalized posterior 
likelihood of the parameter: $\alpha$ for the WDM and the ULA models, respectively. 
The posterior probability of $\alpha$ is displayed in   Figure~\ref{fig:fig7}.
We notice that the marginalized posterior likelihood of the parameter $\alpha$ peaks at $\alpha = -0.1$ (left panel) and $\alpha = -0.05$ (right panel), respectively, for the WDM and the ULA models. The peak values of the parameter $\alpha$ for both the WDM and the ULA models are consistent with $\alpha = 0$ i.e. a redshift-independent background hydrogen-ionizing intensity, in agreement with the findings of \cite{faucher2008flat}. In Figure~\ref{fig:fig9}  we  show  contour
  plots in the spectral index  $\alpha$ and  dark matter masses  plane, which allows
  a further assessment  of the dependence of dark matter mass on  modelling parameters.  In particular, the contour plots capture the degeneracy between
  dark matter masses and the evolution of ionizing radiation, which is
  difficult to determine from theoretical considerations

 To get the posterior likelihoods of the masses for the WDM and the ULA models, we further  marginalize the joint likelihood  of $\alpha$ and the masses over  the parameter $\alpha$. In Figure~\ref{fig:fig6}, we show the posterior probabilities 
of the WDM mass  $m_{\rm wdm}$ (left panel) and the ULA mass  $m_{\rm a}$ (right panel). 
 The posterior probabilities peak at 
$m_{\rm wdm} \simeq  \hbox{a few} \, {\rm keV}$ for the WDM model and at  $m_{\rm a} \simeq  5 \times 10^{-22} \, {\rm eV}$ for the ULA model and are 
flat for larger masses.  As the $\Lambda$CDM model is a limit of these
models for large masses,  the data are  seen to be consistent with the $\Lambda$CDM model. The posterior probabilities
fall for smaller masses which shows that the data are  sensitive to the change in the matter power at small
scales. Figure~\ref{fig:fig6} shows that 1-$\sigma$ (defined 
as the value at which the posterior probability is nearly 0.68) bounds
from our analysis are: $m_{\rm wdm} \gtrsim 0.7 \, \rm keV$ and $m_a \gtrsim 2 \times 10^{-23} \, \rm eV$. 

The discussion in section~\ref{sec:dmmod} allows us to understand these results. For $m_{\rm wdm} \simeq 3 \, \rm keV$, the matter power is smaller than the $\Lambda$CDM model by around 10\% at $k\simeq 9 \, \rm h Mpc^{-1}$. The corresponding scale  for $m_a = 2 \times 10^{-22} \, \rm eV$ is $k\simeq 4 \, \rm h Mpc^{-1}$. For the 1$\sigma$ bounds on  the dark matter masses, the matter power is 
smaller than that for the $\Lambda$CDM model by approximately 10\% at scales $k \simeq 2 \, {\rm h Mpc^{-1}}$.  This suggests that our proposed probe is sensitive to scales comparable to or larger than the Jeans' scale by a factor of 2.
The posterior probability doesn't fall sharply 
for smaller masses which suggests that a range of even larger scales impact the observable. Figure~\ref{fig:fig4} shows that the redshift coverage of the data is partly responsible for this behavior, as the
deviation of the models corresponding to the peak of the posterior probability is less than 3\% at $z\simeq 4.2$, which is the largest redshift in the data. This difference triples for $z \simeq 6$ which means
 the higher  redshift data is a better discriminator between the 
 different models. In the following subsection, we shall compare our 
theoretical predictions with a simulated data set that spans a 
redshift range: $2 \leq z \leq 6.5$. 
 
The Lyman-$\alpha$ data can be analyzed in many different
ways to probe the matter power spectrum at small scales. One way is to model the  the one-dimensional power spectrum of the observed flux field
along the line of sight  using  three-dimensional power spectra at any given redshift (Eq.~(\ref{eq:powspec_a})). Such modelling requires high-resolution hydrodynamical simulations. This method has been successfully applied to probe the matter power spectrum for over two decades now (e.g. \cite{viel2004inferring,palanque2013one,irvsivc2017lyman,chabanier2019one,viel2013warm,viel2008lyman,kobayashi2017lyman}).
Both the WDM and ULA models have been constrained using this approach (\cite{viel2013warm,kobayashi2017lyman,rogers2020strong}). 
The resultant upper bounds on the masses are: $m_{\rm wdm} \gtrsim 3 \, \rm keV$ and $m_a \gtrsim 10^{-21} \, \rm eV$,  which are significantly stronger than our results. These bounds suggest that such probes are sensitive to $k \simeq 10 \, \rm h Mpc^{-1}$, while, as discussed above, our analysis can only probe scales larger than
   $k \simeq 2 \, \rm h Mpc^{-1}$ (see the discussion in section~\ref{sec:dmmod}). As this is partly owing to the redshift coverage of the high-resolution Lyman-$\alpha$ data, we next assess the efficacy of our proposed method using simulated data.

\subsection{Comparison with simulated data}

 We simulate  synthetic data sets for the $\Lambda$CDM model
in the redshift range $2 < z < 6.5$ and at redshift intervals $\Delta z = 0.1$.
The following parameters are used for computing $\tau_{\rm eff}$: $ J = 1.5 \times 10^{-12} \, {\rm s}^{-1}$, $T_0 = 1.3 \times 10^{4} \, {\rm K}$, $\gamma = 1.4$, and $\Delta v = 7.5 \, {\rm km/s}$. The error at each data point  is obtained by drawing from a Gaussian random variable of standard deviation $\sigma = p\tau_{\rm eff}$. Here $p$ is chosen in the range 0.1 and 0.05\footnote{For comparison, $p$ varies in the following range for  the
  existing data sets: $p = 0.15\hbox{--}0.03$ (\cite{faucher2008}) and $p = 0.1\hbox{--}0.02$ (\cite{becker2013}).} 

We use the same priors on the 
parameters: $J$, $T_0$ and $\gamma$, as in the previous section. 
The parameters  $J_0$, $\alpha$, $T_0$ and $\gamma$ are marginalized to obtain the posterior probability 
of masses  $m_{\rm wdm}$ and $m_{\rm a}$ for the WDM and the ULA models respectively. In Figure~\ref{fig:fig8}, we show the posterior probabilities of
the masses for $p = 0.05$. For comparison, the posterior probabilities from
Figure~\ref{fig:fig6} are also shown.

\begin{figure}[h]
\begin{center}
\vskip.2cm \psfrag{k3}[c][c][0.8][0]{redshift $z$}
\psfrag{k1}[c][c][0.8][0]{$\Delta \tau_{\rm avg} (z)$}
\psfrag{k2}[c][c][0.8][0]{$\Delta \tau_{\rm eff} (z)$} 
\psfrag{CDM model}[c][c][0.5][0]{$\Lambda$CDM \quad \quad  } 
\psfrag{WDM model - mwdm = 0.3 keV}[c][c][0.5][0]{\quad \quad \quad $m_{\rm wdm}  = 0.3 \, {\rm keV}$  }
\psfrag{Axion model - ma = 1e-23 keV}[c][c][0.5][0]{\quad \quad \quad \quad $m_{\rm a} = 10^{-23} \, {\rm eV}$  }
\psfrag{WDM model - mwdm = 1.0 keV}[c][c][0.5][0]{\quad \quad \quad $m_{\rm wdm}  = 1.0 \, {\rm keV}$  }
\psfrag{Axion model - ma = 5e-23 keV}[c][c][0.5][0]{\quad \quad $m_{\rm a} = 5 \times 10^{-23} \, {\rm eV}$  }
\centerline{{\includegraphics[scale =0.60]{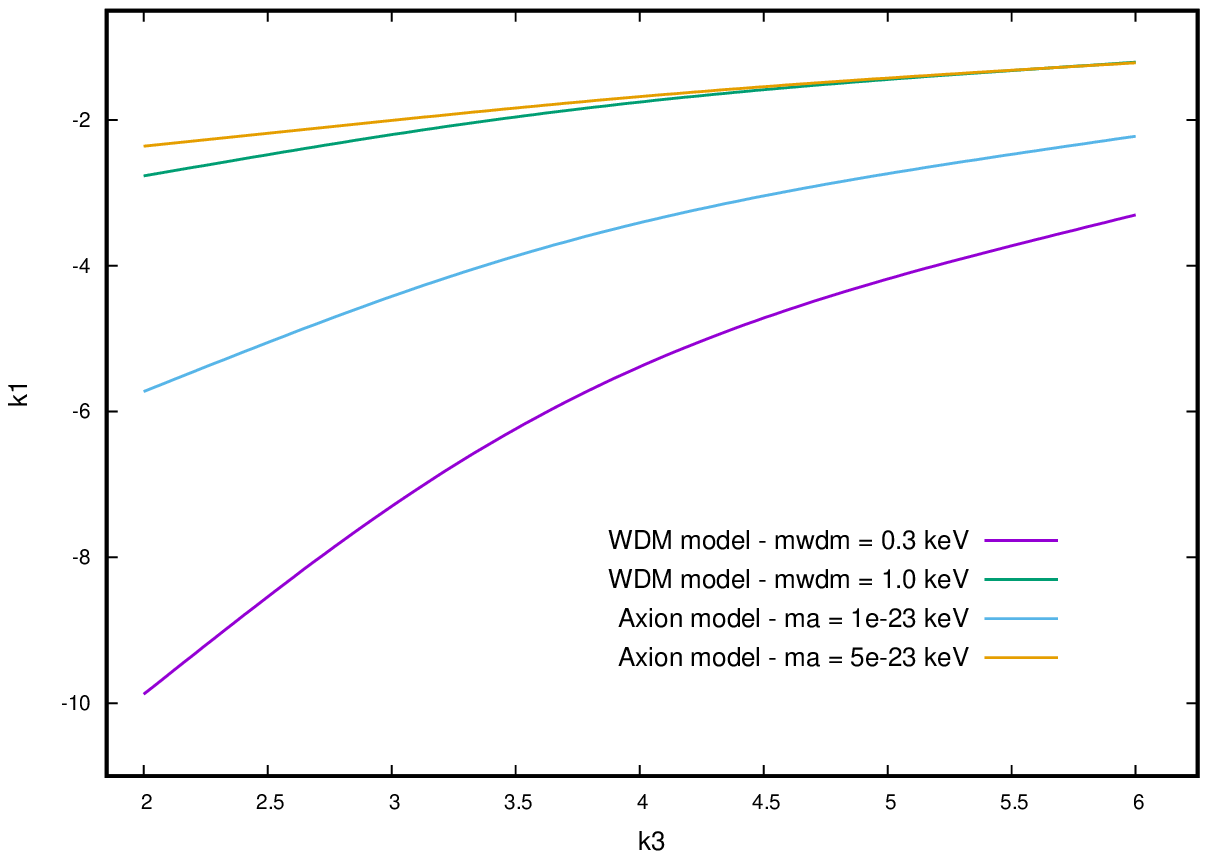}}
  {\includegraphics[scale=0.60]{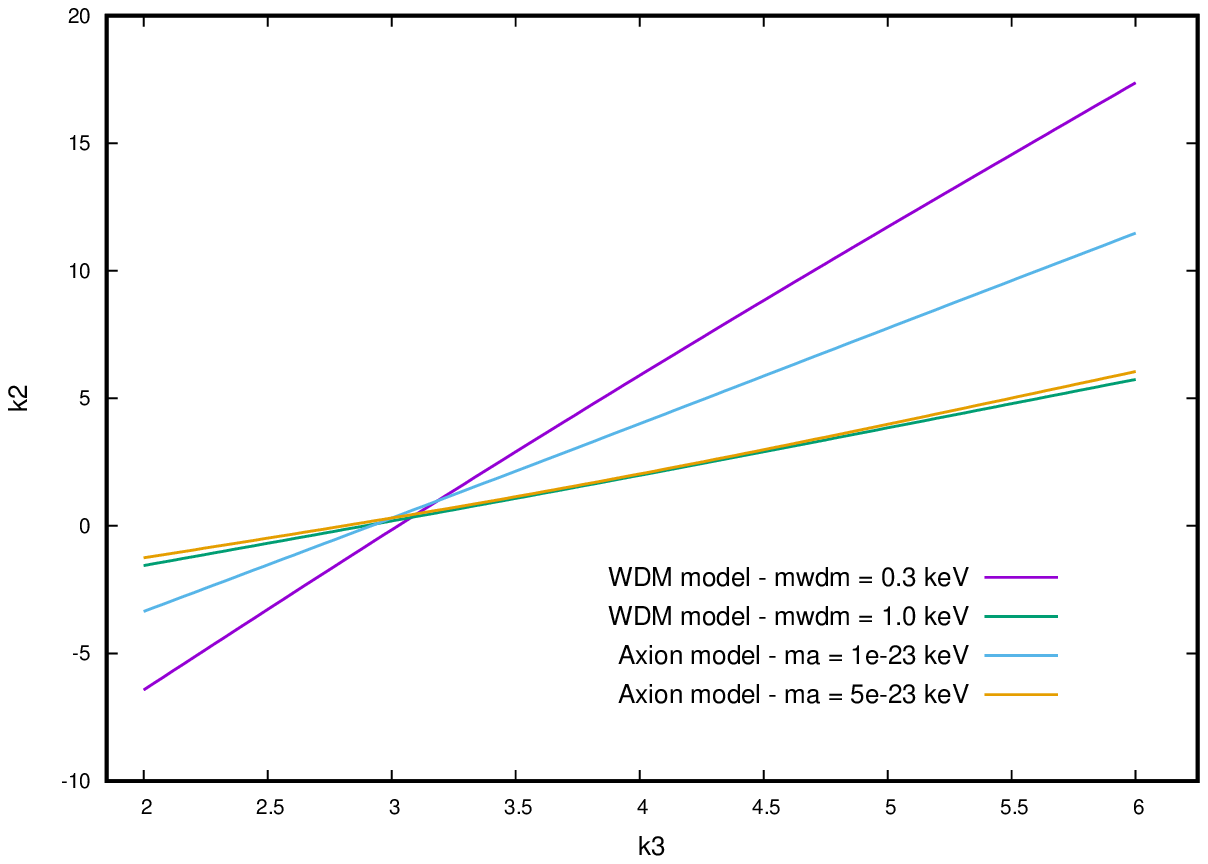}}}
\caption{The  redshift evolution of the relative differences $\Delta \tau_{\rm avg} (z)$ and
  $\Delta \tau_{\rm eff} (z)$, comparing the $\Lambda$CDM model with 
   the WDM  and 
  ULA models, are plotted.}
\label{fig:fig4}
\end{center}
\end{figure}

\begin{figure}[h]
\begin{center}
\vskip.2cm \psfrag{k1}[c][c][1.0][0]{redshift $z$}
\psfrag{k3}[c][c][1.0][0]{$\tau_{\rm eff} (z)$}
\psfrag{CDM model}[c][c][0.75][0]{$\Lambda$CDM \quad \quad  } 
\psfrag{WDM model - mwdm = 0.3 keV}[c][c][0.75][0]{\quad \quad \quad $m_{\rm wdm}  = 0.3 \, {\rm keV}$  }
\psfrag{Axion model - ma = 1e-23 keV}[c][c][0.75][0]{\quad \quad \quad \quad $m_{\rm a} = 10^{-23} \, {\rm eV}$  }
\psfrag{WDM model - mwdm = 1.0 keV}[c][c][0.75][0]{\quad \quad \quad $m_{\rm wdm}  = 1.0 \, {\rm keV}$  }
\psfrag{Axion model - ma = 5e-23 keV}[c][c][0.75][0]{\quad \quad $m_{\rm a} = 5 \times 10^{-23} \, {\rm eV}$  }
\psfrag{FG (2008)}[c][c][0.75][0]{FG (2008) \quad \quad }
\centerline{\includegraphics[scale=0.95]{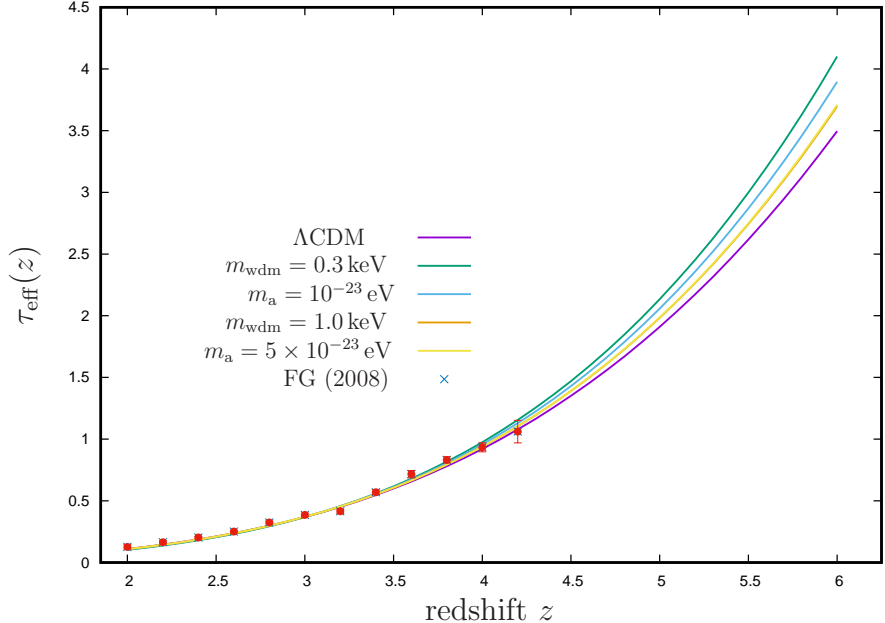}}
\caption{The figure compares the simulated  $\tau_{\rm eff} (z)$, for one set of parameters,  for the $\Lambda$CDM,  the WDM,  and the ULA models, with  the data  \cite{faucher2008}. The other modelling parameters are the same as in Figure~\ref{fig:fig1}. }
\label{fig:fig5}
\end{center}
\end{figure}

\begin{figure}[h]
\begin{center}
\vskip.2cm 
\psfrag{k1}[c][c][0.7][0]{Posterior Probability}
\psfrag{k2}[c][c][0.7][0]{$\alpha$}
\psfrag{a1}[c][c][0.7][0]{}
\psfrag{a2}[c][c][0.7][0]{}
\psfrag{a3}[c][c][0.7][0]{}
\psfrag{b1}[c][c][0.7][0]{$\alpha = -0.1$}
\psfrag{b2}[c][c][0.7][0]{$\alpha = -0.05$}
\centerline{{\includegraphics[scale=0.60]{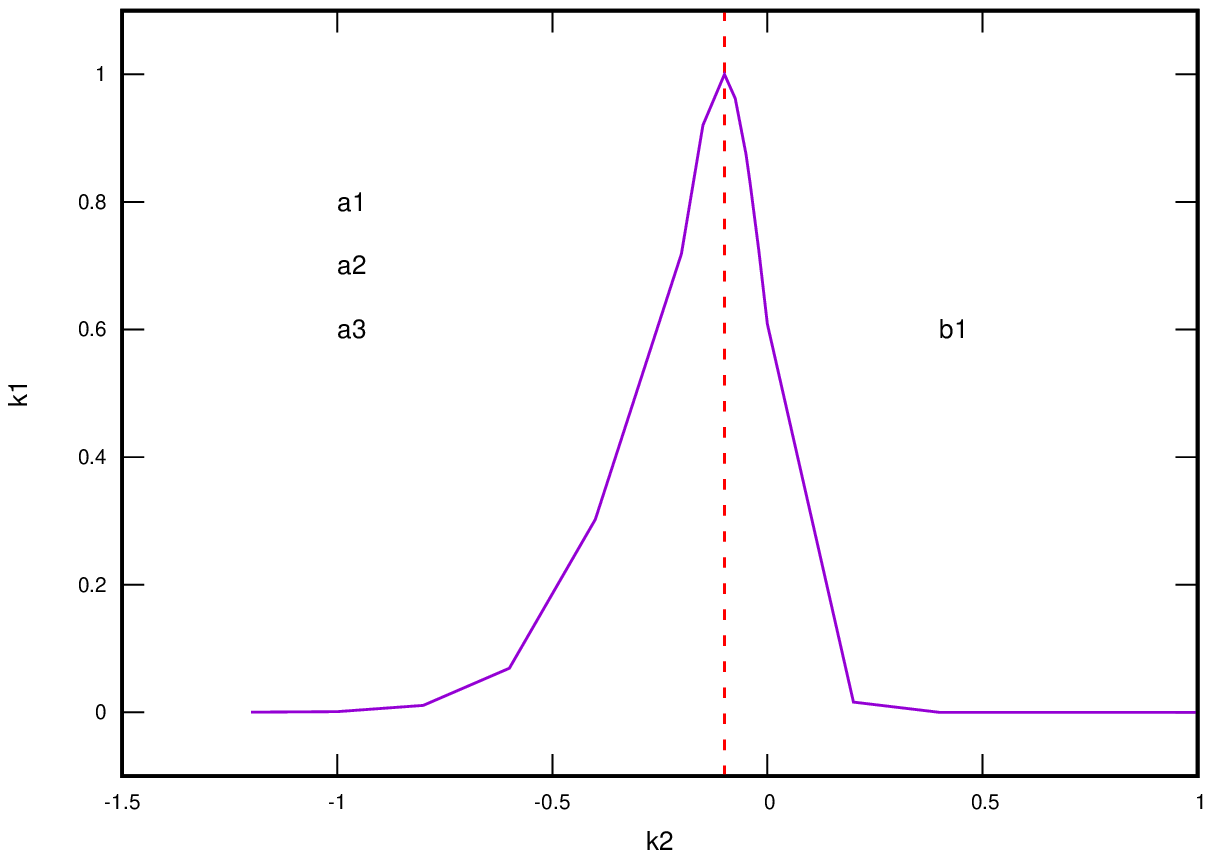}}
  {\includegraphics[scale=0.60]{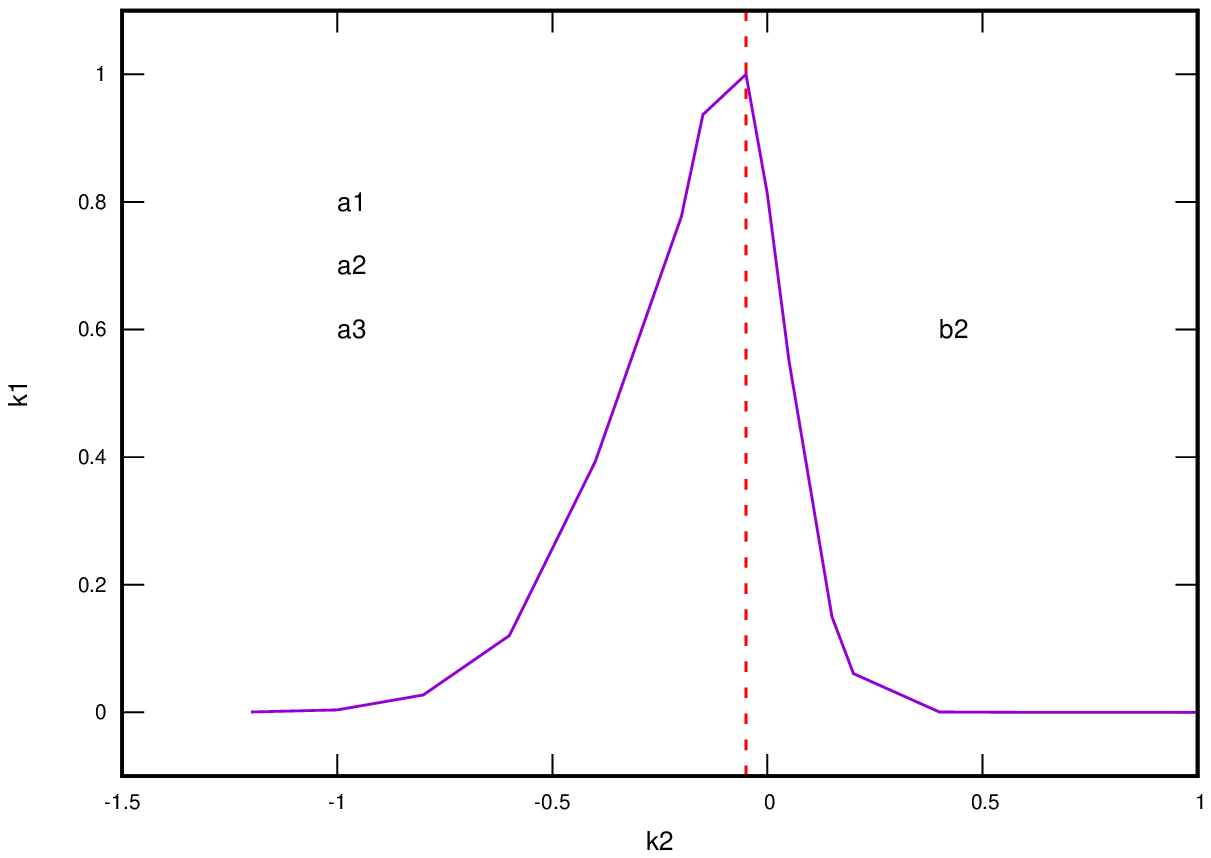}}}
\caption{The figure shows the posterior probabilities of  $\alpha$ (see text for details)  for the WDM (left panel) and the ULA (right panel) models.}
\label{fig:fig7}
\end{center}
\end{figure}

\begin{figure}[h]
\begin{center}
\vskip.2cm 
\psfrag{k2}[c][c][0.7][0]{Posterior Probability}
\psfrag{k1}[c][c][0.7][0]{$m_{\rm wdm}$ (in ${\rm keV}) $}
\psfrag{k3}[c][c][0.7][0]{$m_{\rm a} \, (\times 10^{-22} \, {\rm eV}) $}
\psfrag{a1}[c][c][0.7][0]{}
\psfrag{a2}[c][c][0.7][0]{}
\psfrag{a3}[c][c][0.7][0]{}
\centerline{{\includegraphics[scale=0.60]{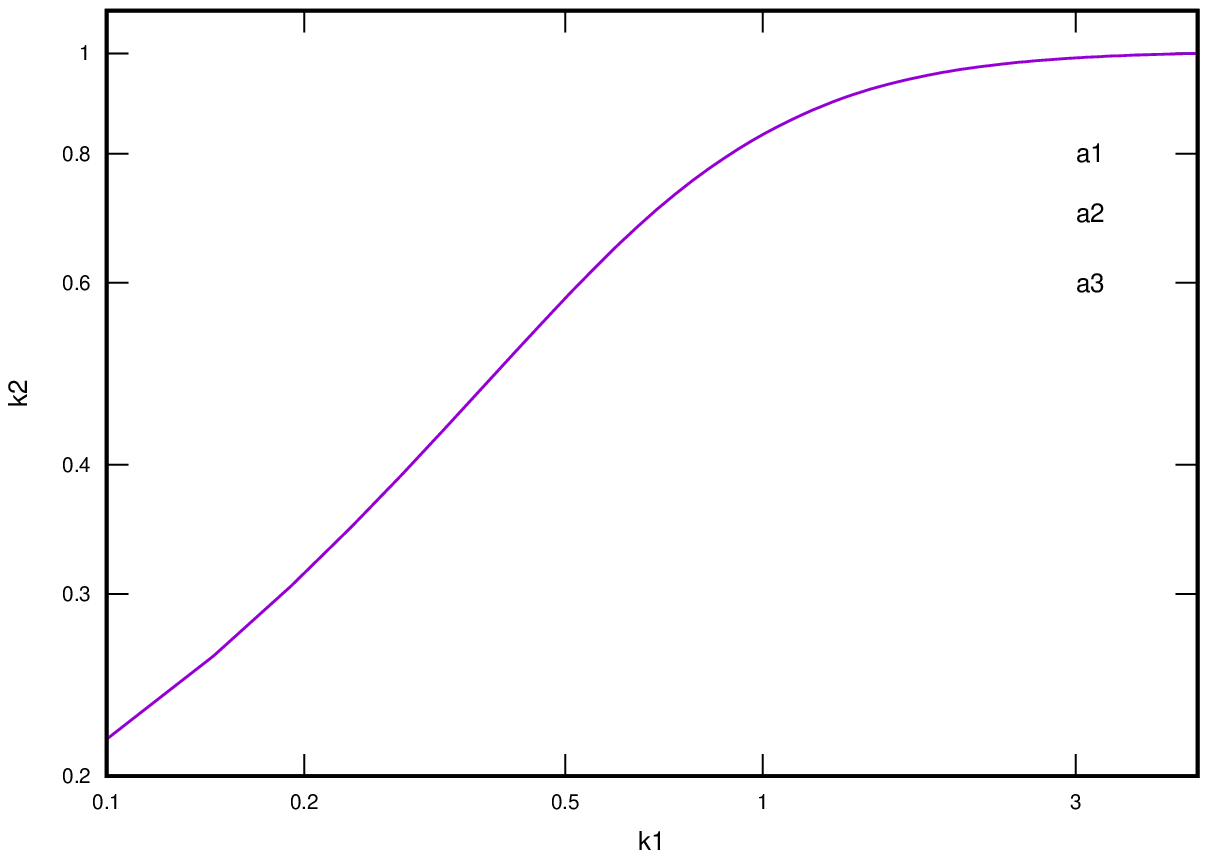}}
  {\includegraphics[scale=0.60]{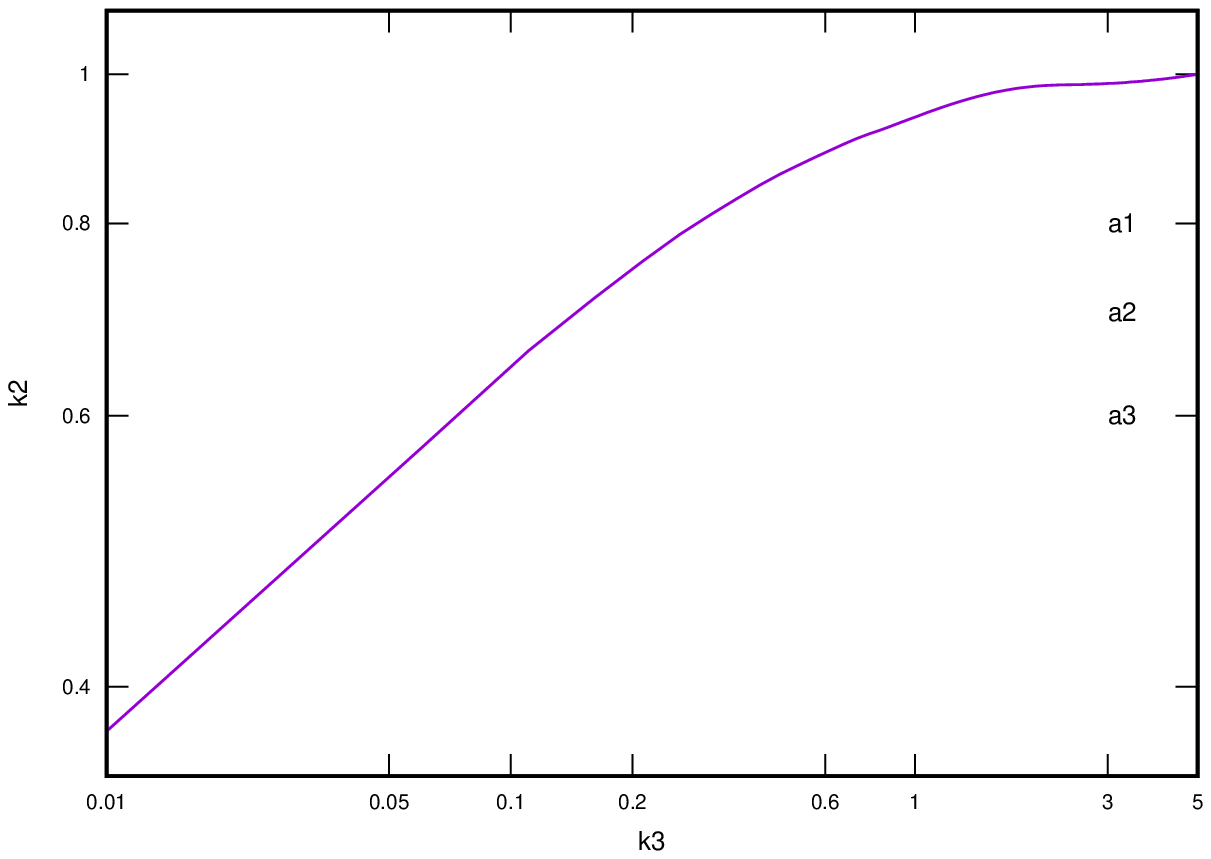}}}
\caption{The posterior probabilities of $m_{\rm wdm}$ and $m_{\rm a}$ for the WDM (left panel) and the ULA (right panel) models are displayed.}
\label{fig:fig6}
\end{center}
\end{figure}

\begin{figure}[h]
\begin{center}
\vskip.2cm 
\psfrag{p1}[c][c][0.45][0]{$1$}
\psfrag{p2}[c][c][0.45][0]{$2.3$}
\psfrag{p3}[c][c][0.45][0]{$6.17$}
\psfrag{p4}[c][c][0.45][0]{$11.8$}
\psfrag{WDM}[c][c][0.7][0]{$m_{\rm wdm}$ (in keV)}
\psfrag{ULA}[c][c][0.7][0]{$m_{\rm a} ( \times 10^{-22} \, {\rm eV})$}
\psfrag{alpha}[c][c][0.7][0]{$\alpha$}
\centerline{{\includegraphics[scale=0.65]{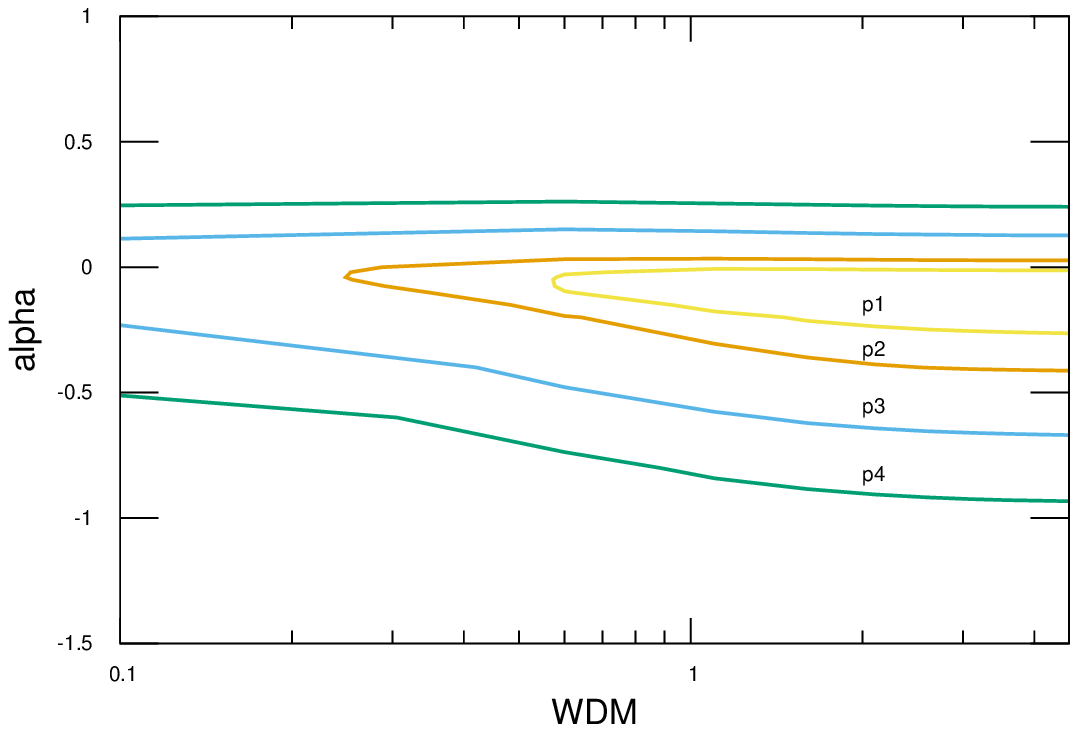}}
  {\includegraphics[scale=0.65]{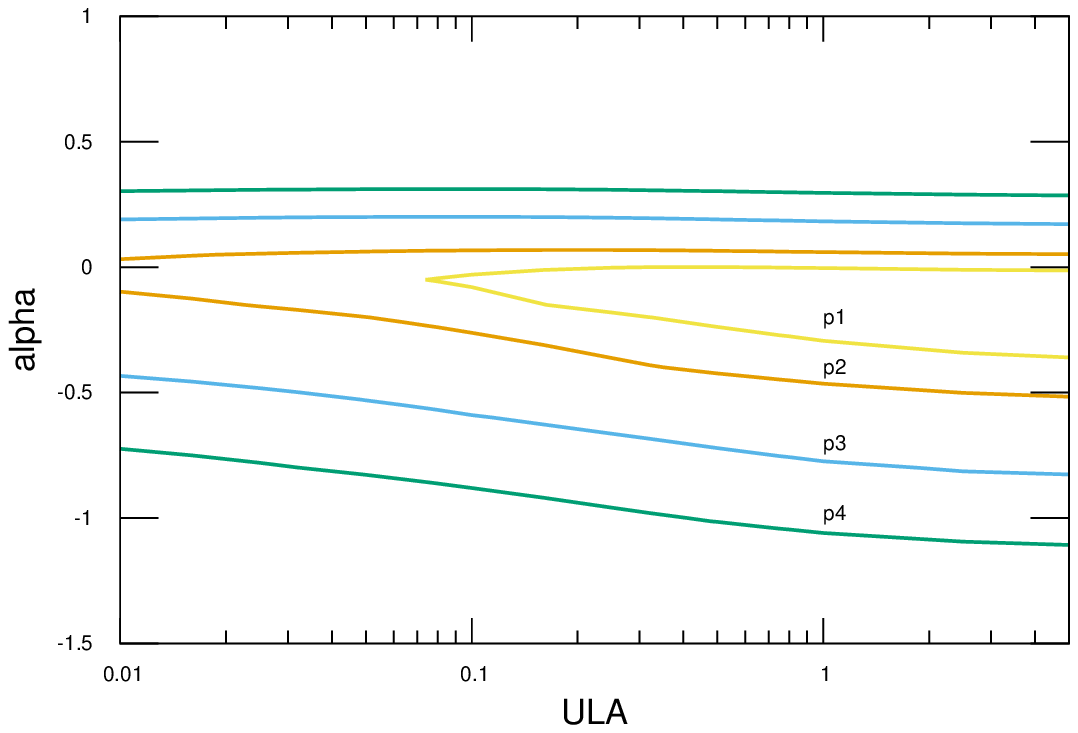}}}
\caption{Figure shows contour plots for parameters: 
$m_{\rm wdm}$ and $\alpha$ for the WDM model (left panel), and $m_{\rm a}$ and 
$\alpha$ for the ULA model (right panel). The legends in the figures 
correspond to $\Delta \chi^2 = 1, 2.3, 6.17$ and $11.8$ (from top to bottom).}
\label{fig:fig9}
\end{center}
\end{figure}

\begin{figure}[h]
\begin{center}
\vskip.2cm 
\psfrag{k1}[c][c][0.7][0]{Posterior Probability}
\psfrag{k2}[c][c][0.7][0]{$m_{\rm wdm}$ (in ${\rm keV}) $}
\psfrag{k3}[c][c][0.7][0]{$m_{\rm a} \, (\times 10^{-22} \, {\rm eV}) $}
\psfrag{FG (2008)}[c][c][0.55][0]{data - FG (2008) \quad \quad \quad \quad }
\psfrag{simulated}[c][c][0.55][0]{data - simulated \quad \quad \quad \quad }
\centerline{{\includegraphics[scale=0.60]{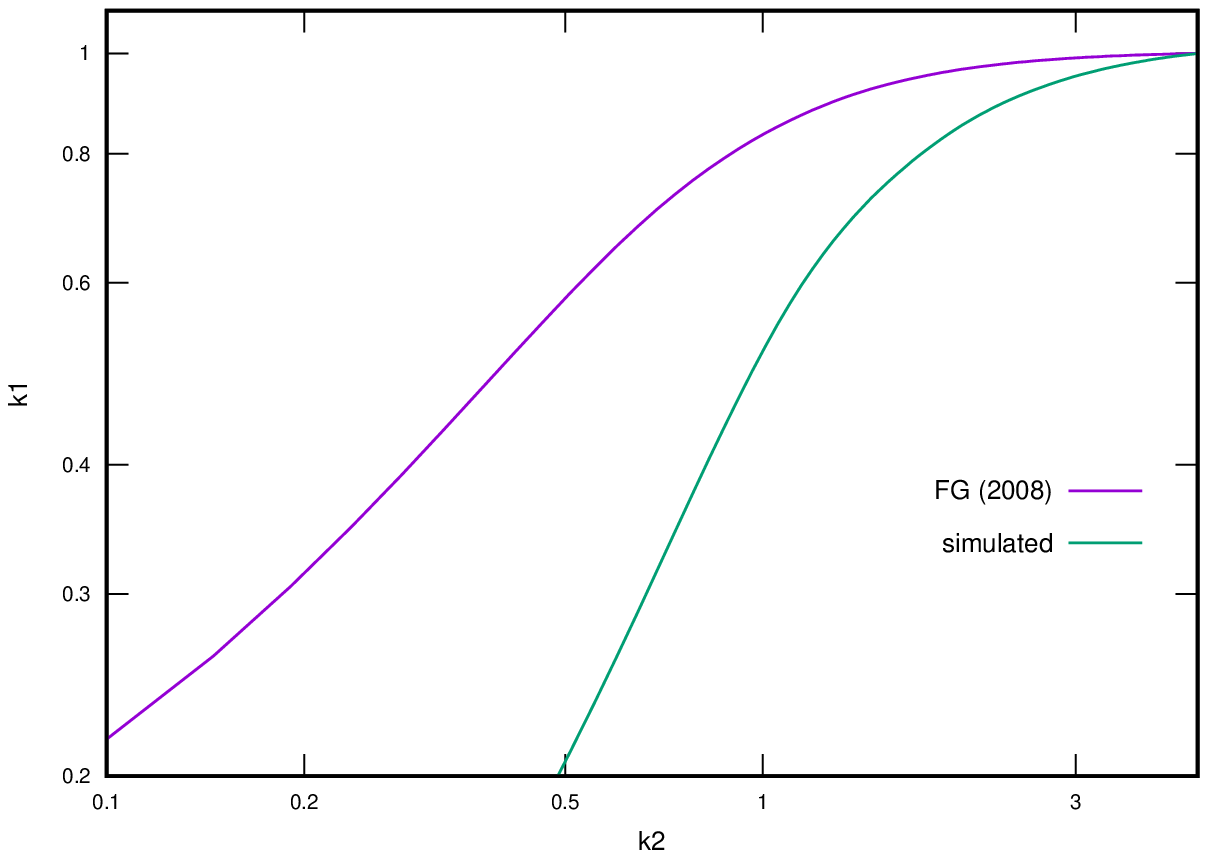}}
  {\includegraphics[scale=0.60]{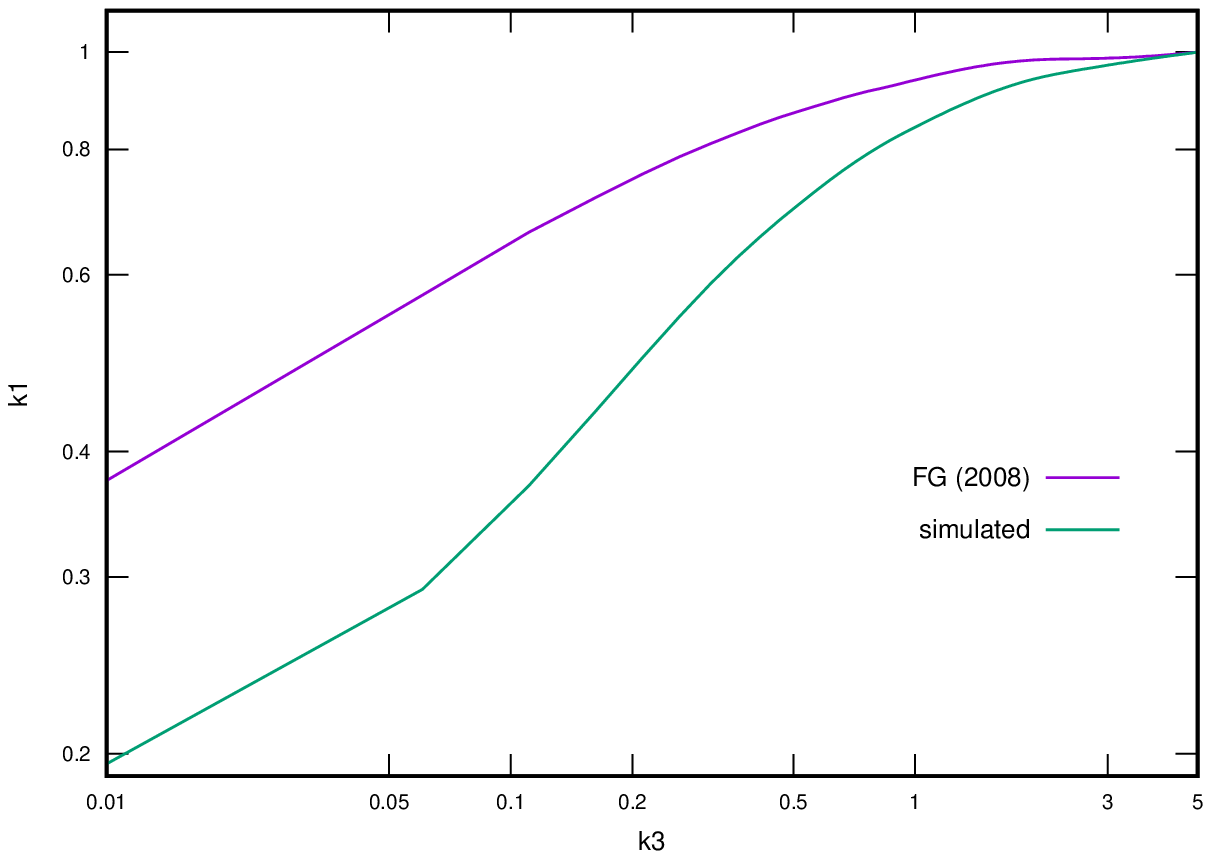}}}
\caption{The Figure shows  posterior probabilities  of dark matter
  masses by comparing the theory with simulated data compatible with the  $\Lambda$CDM model. 
The posterior probabilities from Figure~\ref{fig:fig6} are also shown (magenta curves) for comparison.}
\label{fig:fig8}
\end{center}
\end{figure}

 From Figure~\ref{fig:fig8}, the 1-$\sigma$ forecasts on dark matter masses are: 
$m_{\rm wdm} \gtrsim 1.5 \, \rm keV$ and $m_a \gtrsim 7 \times 10^{-23} \, \rm eV$. 
This is significant improvement over the constraints obtained from the
data set of \cite{faucher2008}. This result  satisfies  our theoretical
expectation (Figure~\ref{fig:fig4}) that the  higher redshift data should provide better constraints. 
We also assess the impact of the    error bar (given by the choice
of $p$) in the simulated data points on the posterior probability. 
For $p = 0.1$, we still find substantial improvement 
but the constraints on WDM mass are found to 
less sensitive to the choice of $p$.

While the comparison of the simulated data sets with the theory bears out the
efficacy of our proposed method, the    forecast bounds   are still weaker
than those obtained from hydrodynamical simulations. As already noted 
(see footnote \ref{fn:jeans}), 
our prescription of implementing the physics close to the 
Jeans' scale results in a sharp cut off in the matter power for 
scales corresponding to $k_J \simeq 5 \, \rm Mpc^{-1}$, which is probably a factor of a few larger than the
scales accessible to hydrodynamical simulations. 

\section{Summary and Conclusions} 
\label{sec:summ}

It is well known that the Lyman-$\alpha$ forest in the redshift range $2 < z < 5$ are an excellent  probe  of  the  matter power spectrum for  scales as small as the  Jeans' scale of the IGM,  $k_J \simeq 5\hbox{--}7 \, {\rm Mpc}^{-1}$.

In this work, we use semi-analytic
modelling of Lyman-$\alpha$ clouds and the redshift evolution of the effective optical depth to  distinguish between
dark matter models whose matter power spectra differ from the $\Lambda$CDM model  at small scales.  This approach  has been shown to be sensitive to the small scale matter power (\cite{2013ApJ...762...15P}). We consider two alternative
dark matter models: ULA and WDM. Both these models yield lower matter power at
small scales as compared to the $\Lambda$CDM model.

We simulate the line-of-sight HI density field corresponding to Lyman-$\alpha$ forest  and as our focus is the probe of small scales, the   simulations resolve the Jeans' scale. Figure~\ref{fig:fig4} allows us to gauge
the impact of the alternative dark matter models of the evolution
of effective optical depth. The effective optical
depth for these model diverges from the prediction of the $\Lambda$CDM model
at higher redshifts. The percentage
difference of these models from  the $\Lambda$CDM model nearly triples in the 
redshift range $4 < z < 6$. Therefore, data at higher redshifts is expected to be 
the principle discriminator between  different dark matter models.

Using Likelihood analysis, we compare the evolution of the effective optical
depth from the simulations with the high-resolution Lyman-$\alpha$ data
\cite{faucher2008} that spans a redshift range: $2 \leq z \leq 4.2$.
The posterior probabilities of the dark matter masses are shown in Figure~\ref{fig:fig6}.
These results yield
the following 1-$\sigma$ bounds on the dark matter masses: 
$m_{\rm wdm} > 0.7\, {\rm keV}$ and 
$m_{\rm a} > 2 \times 10^{-23} \, {\rm eV}$ respectively 
for the WDM and the ULA models.   

We next compare  our theoretical predictions against 
 simulated  $\Lambda$CDM  effective optical depth data sets  over a larger redhift range: $2 \leq z \leq 6.5$. This 
allows us to assess  the efficacy of  high-redshift data as a 
discriminator between the different dark matter models (Figure~\ref{fig:fig4}).
The resulting posterior probabilities are shown in Figure~\ref{fig:fig8}).
The 1-$\sigma$  forecasts  on the dark matter masses are more stringent in this case,  
as expected from the theory, and are given by: 
$m_{\rm wdm} > 1.5 \, {\rm keV}$ and $m_{\rm a} > 7 \times 10^{-23} \, {\rm eV}$. 

The nature of dark matter remains a mystery. One possible way to understand its
nature is to probe cosmological matter power spectrum at progressively smaller scales. 
The future detection  of the  21-cm forest holds the promise of probing even smaller scales 
(\cite{shimabukuro2020constraining}).

\section{Acknowledgment}

 AKS would like to thank RRI for allowing the access to the  
in-house cluster computing facility that was   
used for the computation needed for the analysis in this paper. 
AKS would also like to thank the anonymous referee(s) for 
useful suggestions and critical comments on our paper. 
KLP acknowledges the support of the Department of Atomic Energy, Government of India, 
under project no. RTI4001.

\bibliographystyle{JHEP} 
\bibliography{ref}

\providecommand{\href}[2]{#2}\begingroup\raggedright\begin{thebibliography}{100}

\bibitem{hinshaw2013nine}
G.~Hinshaw, D.~Larson, E.~Komatsu, D.~N. Spergel, C.~Bennett, J.~Dunkley
  et~al., \emph{Nine-year wilkinson microwave anisotropy probe (wmap)
  observations: cosmological parameter results}, {\emph{The Astrophysical
  Journal Supplement Series} {\bfseries 208} (2013) 19}.

\bibitem{Sievers:2013ica}
J.~L. Sievers, R.~A. Hlozek, M.~R. Nolta, V.~Acquaviva, G.~E. Addison, P.~A.
  Ade et~al., \emph{The atacama cosmology telescope: Cosmological parameters
  from three seasons of data}, {\emph{Journal of Cosmology and Astroparticle
  Physics} {\bfseries 2013} (2013) 060}.

\bibitem{aghanim2018planck}
N.~Aghanim, Y.~Akrami, M.~Ashdown, J.~Aumont, C.~Baccigalupi, M.~Ballardini
  et~al., \emph{Planck 2018 results. vi. cosmological parameters}, {\emph{arXiv
  preprint arXiv:1807.06209} (2018) }.

\bibitem{abolfathi2018fourteenth}
B.~Abolfathi, D.~Aguado, G.~Aguilar, C.~A. Prieto, A.~Almeida, T.~T. Ananna
  et~al., \emph{The fourteenth data release of the sloan digital sky survey:
  First spectroscopic data from the extended baryon oscillation spectroscopic
  survey and from the second phase of the apache point observatory galactic
  evolution experiment}, {\emph{The Astrophysical Journal Supplement Series}
  {\bfseries 235} (2018) 42}.

\bibitem{hicken2009improved}
M.~Hicken, W.~M. Wood-Vasey, S.~Blondin, P.~Challis, S.~Jha, P.~L. Kelly
  et~al., \emph{Improved dark energy constraints from~ 100 new cfa supernova
  type ia light curves}, {\emph{The Astrophysical Journal} {\bfseries 700}
  (2009) 1097}.

\bibitem{conley2010supernova}
A.~Conley, J.~Guy, M.~Sullivan, N.~Regnault, P.~Astier, C.~Balland et~al.,
  \emph{Supernova constraints and systematic uncertainties from the first three
  years of the supernova legacy survey}, {\emph{The Astrophysical Journal
  Supplement Series} {\bfseries 192} (2010) 1}.

\bibitem{ho2012clustering}
S.~Ho, A.~Cuesta, H.-J. Seo, R.~De~Putter, A.~J. Ross, M.~White et~al.,
  \emph{Clustering of sloan digital sky survey iii photometric luminous
  galaxies: the measurement, systematics, and cosmological implications},
  {\emph{The Astrophysical Journal} {\bfseries 761} (2012) 14}.

\bibitem{Bartelmann:1999yn}
M.~Bartelmann and P.~Schneider, \emph{Weak gravitational lensing},
  {\emph{Physics Reports} {\bfseries 340} (2001) 291--472}.

\bibitem{Craig:2015xla}
N.~Craig and A.~Katz, \emph{The fraternal wimp miracle}, {\emph{Journal of
  Cosmology and Astroparticle Physics} {\bfseries 2015} (2015) 054}.

\bibitem{Angloher:2011uu}
G.~Angloher, M.~Bauer, I.~Bavykina, A.~Bento, C.~Bucci, C.~Ciemniak et~al.,
  \emph{Results from 730 kg days of the cresst-ii dark matter search},
  {\emph{The European Physical Journal C} {\bfseries 72} (2012) 1971}.

\bibitem{Aprile:2010um}
E.~Aprile, K.~Arisaka, F.~Arneodo, A.~Askin, L.~Baudis, A.~Behrens et~al.,
  \emph{First dark matter results from the xenon100 experiment},
  {\emph{Physical Review Letters} {\bfseries 105} (2010) 131302}.

\bibitem{Ahmed:2010wy}
Z.~Ahmed, D.~Akerib, S.~Arrenberg, C.~Bailey, D.~Balakishiyeva, L.~Baudis
  et~al., \emph{Results from a low-energy analysis of the cdms ii germanium
  data}, {\emph{Physical Review Letters} {\bfseries 106} (2011) 131302}.

\bibitem{Akerib:2013tjd}
D.~S. Akerib, H.~Ara{\'u}jo, X.~Bai, A.~Bailey, J.~Balajthy, S.~Bedikian
  et~al., \emph{First results from the lux dark matter experiment at the
  sanford underground research facility}, {\emph{Physical review letters}
  {\bfseries 112} (2014) 091303}.

\bibitem{Adriani:2010rc}
O.~Adriani, G.~Barbarino, G.~Bazilevskaya, R.~Bellotti, M.~Boezio, E.~Bogomolov
  et~al., \emph{Pamela results on the cosmic-ray antiproton flux from 60 mev to
  180 gev in kinetic energy}, {\emph{Physical Review Letters} {\bfseries 105}
  (2010) 121101}.

\bibitem{FermiLAT:2011ab}
M.~Ackermann, M.~Ajello, A.~Allafort, W.~Atwood, L.~Baldini, G.~Barbiellini
  et~al., \emph{Measurement of separate cosmic-ray electron and positron
  spectra with the fermi large area telescope}, {\emph{Physical Review Letters}
  {\bfseries 108} (2012) 011103}.

\bibitem{Aguilar:2007yf}
M.~Aguilar, J.~Alcaraz, J.~Allaby, B.~Alpat, G.~Ambrosi, H.~Anderhub et~al.,
  \emph{Cosmic-ray positron fraction measurement from 1 to 30 gev with ams-01},
  {\emph{Physics Letters B} {\bfseries 646} (2007) 145--154}.

\bibitem{Goodman:2010yf}
J.~Goodman, M.~Ibe, A.~Rajaraman, W.~Shepherd, T.~M. Tait and H.-B. Yu,
  \emph{Constraints on light majorana dark matter from colliders},
  {\emph{Physics Letters B} {\bfseries 695} (2011) 185--188}.

\bibitem{Fox:2011pm}
P.~J. Fox, R.~Harnik, J.~Kopp and Y.~Tsai, \emph{Missing energy signatures of
  dark matter at the lhc}, {\emph{Physical Review D} {\bfseries 85} (2012)
  056011}.

\bibitem{moore1999dark}
B.~Moore, S.~Ghigna, F.~Governato, G.~Lake, T.~Quinn, J.~Stadel et~al.,
  \emph{Dark matter substructure within galactic halos}, {\emph{The
  Astrophysical Journal Letters} {\bfseries 524} (1999) L19}.

\bibitem{klypin1999missing}
A.~Klypin, A.~V. Kravtsov, O.~Valenzuela and F.~Prada, \emph{Where are the
  missing galactic satellites?}, {\emph{The Astrophysical Journal} {\bfseries
  522} (1999) 82}.

\bibitem{peebles2010nearby}
P.~Peebles and A.~Nusser, \emph{Nearby galaxies as pointers to a better theory
  of cosmic evolution}, {\emph{Nature} {\bfseries 465} (2010) 565--569}.

\bibitem{diemand2007formation}
J.~Diemand, M.~Kuhlen and P.~Madau, \emph{Formation and evolution of galaxy
  dark matter halos and their substructure}, {\emph{The Astrophysical Journal}
  {\bfseries 667} (2007) 859}.

\bibitem{navarro1997universal}
J.~F. Navarro, C.~S. Frenk and S.~D. White, \emph{A universal density profile
  from hierarchical clustering}, {\emph{The Astrophysical Journal} {\bfseries
  490} (1997) 493}.

\bibitem{stadel2009quantifying}
J.~Stadel, D.~Potter, B.~Moore, J.~Diemand, P.~Madau, M.~Zemp et~al.,
  \emph{Quantifying the heart of darkness with ghalo--a multibillion particle
  simulation of a galactic halo}, {\emph{Monthly Notices of the Royal
  Astronomical Society: Letters} {\bfseries 398} (2009) L21--L25}.

\bibitem{Garrison-Kimmel:2014vqa}
S.~Garrison-Kimmel, M.~Boylan-Kolchin, J.~S. Bullock and E.~N. Kirby, \emph{Too
  big to fail in the local group}, {\emph{Monthly Notices of the Royal
  Astronomical Society} {\bfseries 444} (2014) 222--236}.

\bibitem{BoylanKolchin:2011de}
M.~Boylan-Kolchin, J.~S. Bullock and M.~Kaplinghat, \emph{Too big to fail? the
  puzzling darkness of massive milky way subhaloes}, {\emph{Monthly Notices of
  the Royal Astronomical Society: Letters} {\bfseries 415} (2011) L40--L44}.

\bibitem{croft1998}
R.~A. Croft, D.~H. Weinberg, N.~Katz and L.~Hernquist, \emph{Recovery of the
  power spectrum of mass fluctuations from observations of the ly$\alpha$
  forest}, {\emph{The Astrophysical Journal} {\bfseries 495} (1998) 44}.

\bibitem{croft1999}
R.~A. Croft, D.~H. Weinberg, M.~Pettini, L.~Hernquist and N.~Katz, \emph{The
  power spectrum of mass fluctuations measured from the ly$\alpha$ forest at
  redshift z= 2.5}, {\emph{The Astrophysical Journal} {\bfseries 520} (1999)
  1}.

\bibitem{croft2002}
R.~A. Croft, D.~H. Weinberg, M.~Bolte, S.~Burles, L.~Hernquist, N.~Katz et~al.,
  \emph{Toward a precise measurement of matter clustering: Ly$\alpha$ forest
  data at redshifts 2-4}, {\emph{The Astrophysical Journal} {\bfseries 581}
  (2002) 20}.

\bibitem{mcdonald}
P.~Mcdonald, R.~Cen, D.~H. Weinberg, S.~Burles, D.~P. Schneider, D.~J. Schlegel
  et~al., \emph{The linear theory power spectrum from the lyman-$\alpha$ forest
  in the sloan digital sky survey,” astrophys},  in \emph{J. 635}, Citeseer.

\bibitem{rauch1998lyman}
M.~Rauch, \emph{The lyman alpha forest in the spectra of quasistellar objects},
  {\emph{Annual Review of Astronomy and Astrophysics} {\bfseries 36} (1998)
  267--316}.

\bibitem{mandelbaum2003}
R.~Mandelbaum, P.~McDonald, U.~Seljak and R.~Cen, \emph{Precision cosmology
  from the lyman $\alpha$ forest: power spectrum and bispectrum},
  {\emph{Monthly Notices of the Royal Astronomical Society} {\bfseries 344}
  (2003) 776--788}.

\bibitem{viel2004}
M.~Viel, S.~Matarrese, A.~Heavens, M.~Haehnelt, T.-S. Kim, V.~Springel et~al.,
  \emph{The bispectrum of the lyman $\alpha$ forest at z < 2-2.4 from a large
  sample of uves qso absorption spectra (luqas)}, {\emph{Monthly Notices of the
  Royal Astronomical Society} {\bfseries 347} (2004) L26--L30}.

\bibitem{hernquist1996lyman}
L.~Hernquist, N.~Katz, D.~H. Weinberg and J.~Miralda-Escude, \emph{The
  lyman-alpha forest in the cold dark matter model}, {\emph{The Astrophysical
  Journal Letters} {\bfseries 457} (1996) L51}.

\bibitem{mcdonald1999}
P.~McDonald and J.~Miralda-Escud{\'e}, \emph{Measuring the cosmological
  geometry from the ly$\alpha$ forest along parallel lines of sight},
  {\emph{The Astrophysical Journal} {\bfseries 518} (1999) 24}.

\bibitem{lesgourgues2007}
J.~Lesgourgues, R.~Massey, M.~Viel and M.~H{\"a}hnelt, \emph{A combined
  analysis of lyman-alpha forest, 3d weak lensing and wmap year three data},
  tech. rep., 2007.

\bibitem{croft1999cosmological}
R.~A. Croft, W.~Hu and R.~Dave, \emph{Cosmological limits on the neutrino mass
  from the ly $\alpha$ forest}, {\emph{Physical Review Letters} {\bfseries 83}
  (1999) 1092}.

\bibitem{yeche2017}
C.~Y{\`e}che, N.~Palanque-Delabrouille, J.~Baur and H.~D.~M. Des~Bourboux,
  \emph{Constraints on neutrino masses from lyman-alpha forest power spectrum
  with boss and xq-100}, {\emph{Journal of Cosmology and Astroparticle Physics}
  {\bfseries 2017} (2017) 047}.

\bibitem{mcdonald2007dark}
P.~McDonald and D.~J. Eisenstein, \emph{Dark energy and curvature from a future
  baryonic acoustic oscillation survey using the lyman-$\alpha$ forest},
  {\emph{Physical Review D} {\bfseries 76} (2007) 063009}.

\bibitem{slosar2013measurement}
A.~Slosar, V.~Ir{\v{s}}i{\v{c}}, D.~Kirkby, S.~Bailey, T.~Delubac, J.~Rich
  et~al., \emph{Measurement of baryon acoustic oscillations in the
  lyman-$\alpha$ forest fluctuations in boss data release 9}, {\emph{Journal of
  Cosmology and Astroparticle Physics} {\bfseries 2013} (2013) 026}.

\bibitem{delubac2013baryon}
T.~Delubac, J.~Rich, S.~Bailey, A.~Font-Ribera, D.~Kirkby, J.-M. Le~Goff
  et~al., \emph{Baryon acoustic oscillations in the ly$\alpha$ forest of boss
  quasars}, {\emph{Astronomy \& Astrophysics} {\bfseries 552} (2013) A96}.

\bibitem{delubac2015baryon}
T.~Delubac, J.~E. Bautista, J.~Rich, D.~Kirkby, S.~Bailey, A.~Font-Ribera
  et~al., \emph{Baryon acoustic oscillations in the ly$\alpha$ forest of boss
  dr11 quasars}, {\emph{Astronomy \& Astrophysics} {\bfseries 574} (2015) A59}.

\bibitem{faucher2008}
C.-A. Faucher-Giguere, J.~X. Prochaska, A.~Lidz, L.~Hernquist and
  M.~Zaldarriaga, \emph{A direct precision measurement of the intergalactic
  ly$\alpha$ opacity at $2 \leq z \leq 4.2$}, {\emph{The Astrophysical Journal}
  {\bfseries 681} (2008) 831}.

\bibitem{bolton2009evolution}
J.~S. Bolton, S.~P. Oh and S.~R. Furlanetto, \emph{The evolution of the
  ly$\alpha$ forest effective optical depth following he ii reionization},
  {\emph{Monthly Notices of the Royal Astronomical Society} {\bfseries 396}
  (2009) 2405--2418}.

\bibitem{becker2013}
G.~D. Becker, P.~C. Hewett, G.~Worseck and J.~X. Prochaska, \emph{A refined
  measurement of the mean transmitted flux in the ly-$\alpha$ forest over $2 <
  z < 5$ using composite quasar spectra}, {\emph{Monthly Notices of the Royal
  Astronomical Society} {\bfseries 430} (2013) 2067--2081}.

\bibitem{kamble2020measurements}
V.~Kamble, K.~Dawson, H.~d.~M. des Bourboux, J.~Bautista and D.~P. Scheinder,
  \emph{Measurements of effective optical depth in the ly$\alpha$ forest from
  the boss dr12 quasar sample}, {\emph{The Astrophysical Journal} {\bfseries
  892} (2020) 70}.

\bibitem{bidavidson1997}
H.~Bi and A.~F. Davidsen, \emph{Evolution of structure in the intergalactic
  medium and the nature of the ly$\alpha$ forest}, {\emph{The Astrophysical
  Journal} {\bfseries 479} (1997) 523}.

\bibitem{2013ApJ...762...15P}
K.~L. Pandey and S.~K. Sethi, \emph{Probing primordial magnetic fields using
  ly$\alpha$ clouds}, {\emph{The Astrophysical Journal} {\bfseries 762} (2012)
  15}.

\bibitem{zhang2009galactic}
L.~Zhang, J.~Redondo and G.~Sigl, \emph{Galactic signatures of decaying dark
  matter}, {\emph{Journal of Cosmology and Astroparticle Physics} {\bfseries
  2009} (2009) 012}.

\bibitem{boyarsky2008constraints}
A.~Boyarsky, D.~Iakubovskyi, O.~Ruchayskiy and V.~Savchenko, \emph{Constraints
  on decaying dark matter from xmm--newton observations of m31}, {\emph{Monthly
  Notices of the Royal Astronomical Society} {\bfseries 387} (2008)
  1361--1373}.

\bibitem{boyarsky2009realistic}
A.~Boyarsky, J.~Lesgourgues, O.~Ruchayskiy and M.~Viel, \emph{Realistic sterile
  neutrino dark matter with kev mass does not contradict cosmological bounds},
  {\emph{Physical review letters} {\bfseries 102} (2009) 201304}.

\bibitem{seljak2006can}
U.~Seljak, A.~Makarov, P.~McDonald and H.~Trac, \emph{Can sterile neutrinos be
  the dark matter?}, {\emph{Physical Review Letters} {\bfseries 97} (2006)
  191303}.

\bibitem{smith2011}
R.~E. Smith and K.~Markovic, \emph{Testing the warm dark matter paradigm with
  large-scale structures}, {\emph{Physical Review D} {\bfseries 84} (2011)
  063507}.

\bibitem{ellis1984supersymmetric}
J.~Ellis, J.~Hagelin, D.~Nanopoulos, K.~Olive and M.~Srednicki,
  \emph{Supersymmetric relics from the big bang},  1984.

\bibitem{dodelson1994sterile}
S.~Dodelson and L.~M. Widrow, \emph{Sterile neutrinos as dark matter},
  {\emph{Physical Review Letters} {\bfseries 72} (1994) 17}.

\bibitem{arvanitaki2010string}
A.~Arvanitaki, S.~Dimopoulos, S.~Dubovsky, N.~Kaloper and J.~March-Russell,
  \emph{String axiverse}, {\emph{Physical Review D} {\bfseries 81} (2010)
  123530}.

\bibitem{frieman1995cosmology}
J.~A. Frieman, C.~T. Hill, A.~Stebbins and I.~Waga, \emph{Cosmology with
  ultralight pseudo nambu-goldstone bosons}, {\emph{Physical Review Letters}
  {\bfseries 75} (1995) 2077}.

\bibitem{coble1997dynamical}
K.~Coble, S.~Dodelson and J.~A. Frieman, \emph{Dynamical $\lambda$ models of
  structure formation}, {\emph{Physical Review D} {\bfseries 55} (1997) 1851}.

\bibitem{hu2000fuzzy}
W.~Hu, R.~Barkana and A.~Gruzinov, \emph{Fuzzy cold dark matter: the wave
  properties of ultralight particles}, {\emph{Physical Review Letters}
  {\bfseries 85} (2000) 1158}.

\bibitem{marsh2010ultralight}
D.~J. Marsh and P.~G. Ferreira, \emph{Ultralight scalar fields and the growth
  of structure in the universe}, {\emph{Physical Review D} {\bfseries 82}
  (2010) 103528}.

\bibitem{park2012axion}
C.-G. Park, J.-c. Hwang and H.~Noh, \emph{Axion as a cold dark matter
  candidate: low-mass case}, {\emph{Physical Review D} {\bfseries 86} (2012)
  083535}.

\bibitem{kobayashi2017lyman}
T.~Kobayashi, R.~Murgia, A.~De~Simone, V.~Ir{\v{s}}i{\v{c}} and M.~Viel,
  \emph{Lyman-$\alpha$ constraints on ultralight scalar dark matter:
  Implications for the early and late universe}, {\emph{Physical Review D}
  {\bfseries 96} (2017) 123514}.

\bibitem{viel2013warm}
M.~Viel, G.~D. Becker, J.~S. Bolton and M.~G. Haehnelt, \emph{Warm dark matter
  as a solution to the small scale crisis: New constraints from high redshift
  lyman-$\alpha$ forest data}, {\emph{Physical Review D} {\bfseries 88} (2013)
  043502}.

\bibitem{Polisensky:2010rw}
E.~Polisensky and M.~Ricotti, \emph{Constraints on the dark matter particle
  mass from the number of milky way satellites}, {\emph{Physical Review D}
  {\bfseries 83} (2011) 043506}.

\bibitem{Anderhalden:2012qt}
D.~Anderhalden, J.~Diemand, G.~Bertone, A.~V. Maccio and A.~Schneider,
  \emph{The galactic halo in mixed dark matter cosmologies}, {\emph{Journal of
  Cosmology and Astroparticle Physics} {\bfseries 2012} (2012) 047}.

\bibitem{Lovell:2011rd}
M.~R. Lovell, V.~Eke, C.~S. Frenk, L.~Gao, A.~Jenkins, T.~Theuns et~al.,
  \emph{The haloes of bright satellite galaxies in a warm dark matter
  universe}, {\emph{Monthly Notices of the Royal Astronomical Society}
  {\bfseries 420} (2012) 2318--2324}.

\bibitem{Maccio:2012qf}
A.~V. Macci{\`o}, S.~Paduroiu, D.~Anderhalden, A.~Schneider and B.~Moore,
  \emph{Cores in warm dark matter haloes: a catch 22 problem}, {\emph{Monthly
  Notices of the Royal Astronomical Society} {\bfseries 424} (2012)
  1105--1112}.

\bibitem{Schneider:2011yu}
A.~Schneider, R.~E. Smith, A.~V. Macci{\`o} and B.~Moore, \emph{Non-linear
  evolution of cosmological structures in warm dark matter models},
  {\emph{Monthly Notices of the Royal Astronomical Society} {\bfseries 424}
  (2012) 684--698}.

\bibitem{Baur:2015jsy}
J.~Baur, N.~Palanque-Delabrouille, C.~Y{\`e}che, C.~Magneville and M.~Viel,
  \emph{Lyman-alpha forests cool warm dark matter}, {\emph{Journal of Cosmology
  and Astroparticle Physics} {\bfseries 2016} (2016) 012}.

\bibitem{Marsh:2015wka}
D.~J. Marsh and A.-R. Pop, \emph{Axion dark matter, solitons and the cusp--core
  problem}, {\emph{Monthly Notices of the Royal Astronomical Society}
  {\bfseries 451} (2015) 2479--2492}.

\bibitem{Marsh:2015xka}
D.~J. Marsh, \emph{Axion cosmology}, {\emph{Physics Reports} {\bfseries 643}
  (2016) 1--79}.

\bibitem{Hui:2016ltb}
L.~Hui, J.~P. Ostriker, S.~Tremaine and E.~Witten, \emph{On the hypothesis that
  cosmological dark matter is composed of ultra-light bosons}, {\emph{arXiv
  preprint arXiv:1610.08297} (2016) }.

\bibitem{2016JCAP...04..012S}
A.~Sarkar, R.~Mondal, S.~Das, S.~K. Sethi, S.~Bharadwaj and D.~J. Marsh,
  \emph{The effects of the small-scale dm power on the cosmological neutral
  hydrogen (hi) distribution at high redshifts}, {\emph{Journal of Cosmology
  and Astroparticle Physics} {\bfseries 2016} (2016) 012}.

\bibitem{2017JCAP...07..012S}
A.~Sarkar, S.~K. Sethi and S.~Das, \emph{The effects of the small-scale
  behaviour of dark matter power spectrum on cmb spectral distortion},
  {\emph{Journal of Cosmology and Astroparticle Physics} {\bfseries 2017}
  (2017) 012}.

\bibitem{hlozek2015}
R.~Hlozek, D.~Grin, D.~J. Marsh and P.~G. Ferreira, \emph{A search for
  ultralight axions using precision cosmological data}, {\emph{Physical Review
  D} {\bfseries 91} (2015) 103512}.

\bibitem{rogers2020strong}
K.~K. Rogers and H.~V. Peiris, \emph{Strong bound on canonical ultra-light
  axion dark matter from the lyman-alpha forest}, {\emph{arXiv preprint
  arXiv:2007.12705} (2020) }.

\bibitem{palanque2013one}
N.~Palanque-Delabrouille, C.~Y{\`e}che, A.~Borde, J.-M. Le~Goff, G.~Rossi,
  M.~Viel et~al., \emph{The one-dimensional ly$\alpha$ forest power spectrum
  from boss}, {\emph{Astronomy \& Astrophysics} {\bfseries 559} (2013) A85}.

\bibitem{irvsivc2017lyman}
V.~Ir{\v{s}}i{\v{c}}, M.~Viel, T.~A. Berg, V.~D'Odorico, M.~G. Haehnelt,
  S.~Cristiani et~al., \emph{The lyman $\alpha$ forest power spectrum from the
  xq-100 legacy survey}, {\emph{Monthly Notices of the Royal Astronomical
  Society} {\bfseries 466} (2017) 4332--4345}.

\bibitem{chabanier2019one}
S.~Chabanier, N.~Palanque-Delabrouille, C.~Y{\`e}che, J.-M. Le~Goff,
  E.~Armengaud, J.~Bautista et~al., \emph{The one-dimensional power spectrum
  from the sdss dr14 ly$\alpha$ forests}, {\emph{Journal of Cosmology and
  Astroparticle Physics} {\bfseries 2019} (2019) 017}.

\bibitem{bond1980massive}
J.~R. Bond, G.~Efstathiou and J.~Silk, \emph{Massive neutrinos and the
  large-scale structure of the universe}, {\emph{Physical Review Letters}
  {\bfseries 45} (1980) }.

\bibitem{viel2005constraining}
M.~Viel, J.~Lesgourgues, M.~G. Haehnelt, S.~Matarrese and A.~Riotto,
  \emph{Constraining warm dark matter candidates including sterile neutrinos
  and light gravitinos with wmap and the lyman-$\alpha$ forest},
  {\emph{Physical Review D} {\bfseries 71} (2005) 063534}.

\bibitem{smith2011testing}
R.~E. Smith and K.~Markovic, \emph{Testing the warm dark matter paradigm with
  large-scale structures}, {\emph{Physical Review D} {\bfseries 84} (2011)
  063507}.

\bibitem{seigar2015cold}
M.~S. Seigar, \emph{Cold dark matter, hot dark matter, and their alternatives},
   in \emph{Dark Matter in the Universe}.
\newblock Morgan \& Claypool Publishers, 2015.

\bibitem{kang2020warm}
X.~Kang, \emph{Warm dark matter model with a few kev mass is bad for the
  too-big-to-fail problem}, {\emph{Monthly Notices of the Royal Astronomical
  Society} {\bfseries 491} (2020) 2520--2535}.

\bibitem{zentner2003halo}
A.~R. Zentner and J.~S. Bullock, \emph{Halo substructure and the power
  spectrum}, {\emph{The Astrophysical Journal} {\bfseries 598} (2003) 49}.

\bibitem{bode2001halo}
P.~Bode, J.~P. Ostriker and N.~Turok, \emph{Halo formation in warm dark matter
  models}, {\emph{The Astrophysical Journal} {\bfseries 556} (2001) 93}.

\bibitem{markovic2011constraining}
K.~Markovic, S.~Bridle, A.~Slosar and J.~Weller, \emph{Constraining warm dark
  matter with cosmic shear power spectra}, {\emph{Journal of Cosmology and
  Astroparticle Physics} {\bfseries 2011} (2011) 022}.

\bibitem{Arvanitaki:2009fg}
A.~Arvanitaki, S.~Dimopoulos, S.~Dubovsky, N.~Kaloper and J.~March-Russell,
  \emph{String axiverse}, {\emph{Physical Review D} {\bfseries 81} (2010)
  123530}.

\bibitem{Marsh:2010wq}
D.~J. Marsh and P.~G. Ferreira, \emph{Ultralight scalar fields and the growth
  of structure in the universe}, {\emph{Physical Review D} {\bfseries 82}
  (2010) 103528}.

\bibitem{Hu:2000ke}
W.~Hu, R.~Barkana and A.~Gruzinov, \emph{Fuzzy cold dark matter: the wave
  properties of ultralight particles}, {\emph{Physical Review Letters}
  {\bfseries 85} (2000) 1158}.

\bibitem{Amendola:2005ad}
L.~Amendola and R.~Barbieri, \emph{Dark matter from an ultra-light
  pseudo-goldsone-boson}, {\emph{Physics Letters B} {\bfseries 642} (2006)
  192--196}.

\bibitem{choudhury2001semianalytic}
T.~R. Choudhury, R.~Srianand and T.~Padmanabhan, \emph{Semianalytic approach to
  understanding the distribution of neutral hydrogen in the universe:
  comparison of simulations with observations}, {\emph{The Astrophysical
  Journal} {\bfseries 559} (2001) 29}.

\bibitem{hui1997}
L.~Hui and N.~Y. Gnedin, \emph{Equation of state of the photoionized
  intergalactic medium}, {\emph{Monthly Notices of the Royal Astronomical
  Society} {\bfseries 292} (1997) 27--42}.

\bibitem{prochaska1999}
J.~X. Prochaska and A.~M. Wolfe, \emph{Chemical abundances of the damped
  ly$\alpha$ systems at z> 1.5}, {\emph{The Astrophysical Journal Supplement
  Series} {\bfseries 121} (1999) 369}.

\bibitem{prochaska2007}
J.~X. Prochaska, A.~M. Wolfe, J.~C. Howk, E.~Gawiser, S.~M. Burles and
  J.~Cooke, \emph{The ucsd/keck damped ly$\alpha$ abundance database: a decade
  of high-resolution spectroscopy}, {\emph{The Astrophysical Journal Supplement
  Series} {\bfseries 171} (2007) 29}.

\bibitem{meara2007}
J.~M. O’meara, J.~X. Prochaska, S.~Burles, G.~Prochter, R.~A. Bernstein and
  K.~M. Burgess, \emph{The keck+ magellan survey for lyman limit absorption. i.
  the frequency distribution of super lyman limit systems}, {\emph{The
  Astrophysical Journal} {\bfseries 656} (2007) 666}.

\bibitem{schneider2010}
D.~P. Schneider, G.~T. Richards, P.~B. Hall, M.~A. Strauss, S.~F. Anderson,
  T.~A. Boroson et~al., \emph{The sloan digital sky survey quasar catalog. v.
  seventh data release}, {\emph{The Astronomical Journal} {\bfseries 139}
  (2010) 2360}.

\bibitem{mcquinn2016evolution}
M.~McQuinn, \emph{The evolution of the intergalactic medium}, {\emph{Annual
  Review of Astronomy and Astrophysics} {\bfseries 54} (2016) 313--362}.

\bibitem{plante2018helium}
P.~La~Plante, H.~Trac, R.~Croft and R.~Cen, \emph{Helium reionization
  simulations. iii. the helium ly$\alpha$ forest}, {\emph{The Astrophysical
  Journal} {\bfseries 868} (2018) 106}.

\bibitem{draine2010physics}
B.~T. Draine, \emph{Physics of the interstellar and intergalactic medium},
  vol.~19.
\newblock Princeton University Press, 2010.

\bibitem{lukic2015lyman}
Z.~Luki{\'c}, C.~W. Stark, P.~Nugent, M.~White, A.~A. Meiksin and A.~Almgren,
  \emph{The lyman $\alpha$ forest in optically thin hydrodynamical
  simulations}, {\emph{Monthly Notices of the Royal Astronomical Society}
  {\bfseries 446} (2015) 3697--3724}.

\bibitem{faucher2008flat}
C.-A. Faucher-Giguere, A.~Lidz, L.~Hernquist and M.~Zaldarriaga, \emph{A flat
  photoionization rate at 2≤ z≤ 4.2: Evidence for a stellar-dominated uv
  background and against a decline of cosmic star formation beyond z ~ 3},
  {\emph{The Astrophysical Journal Letters} {\bfseries 682} (2008) L9}.

\bibitem{viel2004inferring}
M.~Viel, M.~G. Haehnelt and V.~Springel, \emph{Inferring the dark matter power
  spectrum from the lyman $\alpha$ forest in high-resolution qso absorption
  spectra}, {\emph{Monthly Notices of the Royal Astronomical Society}
  {\bfseries 354} (2004) 684--694}.

\bibitem{viel2008lyman}
M.~Viel, \emph{The lyman-$\alpha$ forest as a probe of the coldness of dark
  matter},  in \emph{IFAE 2007}, pp.~255--260.
\newblock Springer, 2008.

\bibitem{shimabukuro2020constraining}
H.~Shimabukuro, K.~Ichiki and K.~Kadota, \emph{Constraining the nature of ultra
  light dark matter particles with the 21 cm forest}, {\emph{Physical Review D}
  {\bfseries 101} (2020) 043516}.

\end{thebibliography}\endgroup

\end{document}